\documentclass[twoside,11pt]{article}

\usepackage{blindtext}

\usepackage{bbm}
\usepackage[ruled,vlined]{algorithm2e}

\usepackage{jmlr2e}

\usepackage{amsmath}
\usepackage{mathtools}   
\usepackage{bm}
\usepackage{mathrsfs}

\usepackage{xcolor}
\usepackage{booktabs}
\usepackage{multirow}
\usepackage{wrapfig}
\usepackage{framed}
\usepackage{suffix}
\usepackage{qtree}
\usepackage{empheq}
\usepackage[most]{tcolorbox}
\usepackage{lastpage}    
\usepackage[inline]{enumitem}

\hypersetup{colorlinks=true, linkcolor=blue, citecolor=blue, urlcolor=blue}

\IfFileExists{math-commands.tex}{%
\def\PG{P{\'o}lya-Gamma}

\newcommand{\diag}{{\rm diag}}

\renewcommand{\det}[1]{\left| #1 \right|}

\renewcommand{\b}{{\bf b}}

\newcommand{\e}{{\rm e}} 

\newcommand{\s}{{\bf s}}

\newcommand{\w}{{\bf w}}
\newcommand{\x}{{\bf x}}
\newcommand{\y}{{\bf y}}
\newcommand{\z}{{\bf z}}

\newcommand{\E}{{\mathbb E}}
\newcommand{\F}{{\bf F}}

\newcommand{\I}{{\bf I}}

\renewcommand{\S}{{\bf S}}

\newcommand{\X}{{\bf X}}
\newcommand{\Y}{{\bf Y}}

\newcommand{\bepsilon}{\boldsymbol{\epsilon}}

\newcommand{\0}{{\bf 0}}

\newcommand{\ben}{\begin{enumerate}}
\newcommand{\een}{\end{enumerate}}
\newcommand{\beq}{\begin{equation}}
\newcommand{\eeq}{\end{equation}}

\newcommand{\abs}[1]{\lvert#1\rvert}

\newcommand{\NormRV}{\mathcal{N}}
\newcommand{\InvGammaRV}{\mathcal{IG}}

\newcommand{\ExpRV}{\mathcal{E}xp}

\usepackage{booktabs,array}

\newcount\rowc

}{}

\def\PG{P{\'o}lya-Gamma}

\jmlrheading{23}{2025}{1--\pageref{LastPage}}{10/23; Revised xx/xx}{xx/xx}{25-000}{Jyotishka Datta, Soham Ghosh, and Nicholas G.\ Polson}

\ShortHeadings{Horseshoe-type Priors for ICA}{Datta, Ghosh and Polson}
\firstpageno{1}

\begin{document}

\title{Bayesian ICA with super-Gaussian Source Priors}

\author{\name Jyotishka Datta \email jyotishka@vt.edu \\
       \addr Department of Statistics, Virginia Tech
       \AND
       \name Soham Ghosh \email sghosh39@wisc.edu \\
       \addr Department of Statistics, University of Wisconsin--Madison
       \AND
       \name Nicholas G.\ Polson \email ngp@chicagobooth.edu \\
       \addr Booth School of Business, University of Chicago}

\editor{TBD}

\maketitle

\begin{abstract}
\noindent 
Independent Component Analysis (ICA) plays a central role in modern machine learning as a flexible framework for feature extraction. We introduce a horseshoe-type prior with a latent \PG{} scale-mixture representation, yielding scalable algorithms for both point estimation via expectation–maximization (EM) and full posterior inference via Markov chain Monte Carlo (MCMC). This hierarchical formulation unifies several previously disparate estimation strategies within a single Bayesian framework. We also establish the first theoretical guarantees for hierarchical Bayesian ICA, including posterior contraction and local asymptotic normality results for the unmixing matrix. Comprehensive simulation studies demonstrate that our methods perform competitively with widely used ICA tools. We further discuss implementation of conditional posteriors, envelope-based optimization, and possible extensions to flow-based architectures for nonlinear feature extraction and deep learning. Finally, we outline several promising directions for future work.
\end{abstract}




\newpage
\section{Introduction}\label{sec:intro}
Linear independent component analysis (ICA) \footnote{Some authors e.g., \citet{dinh2014nice} use the acronym ICE, for Independent Component Estimation.} is central to blind source separation and has many fields of application, e.g., signal processing, medical imaging, machine learning and many others.  ICA can be viewed as a feature extraction problem where a random vector $\bm{x}$ with coordinates that  admit a representation as linear combination of independent latent variables $\bm{s}$ with an unknown mixing matrix $\bm{A}$, namely $ \bm{x} = \bm{A} \bm{s} $.
One needs to estimate $\bm{A}$ or the unmixing matrix $\bm{W} = \bm{A}^{-1}$ \citep{comon1991blind, comon1994independent, bell1995information, hyvarinen1999nonlinear}. ICA can be viewed of as a refinement or an extension of principle component analysis (PCA) or factor analysis, in that the PCA only requires uncorrelated components, not independent. 

Feature engineering (a.k.a. nonlinear factor analysis) is a fundamental problem in many modern day machine learning applications ranging from medical imaging to signal processing. A key aspect of feature engineering is decomposing high-dimensional data into independent latent factors. \citet{bhadra2024merging} presents a statistical view of high-dimensional deep learning. In this formulation, the output $\bm{Y}$ and input $\bm{X}$ connected via a statistical model: 
\[
    P(\bm{Y} \mid \bm{W}, \bm{X}) = P(\bm{Y} \mid \bm{a}), \; \bm{a} = \bm{W} \bm{X}, 
\]
where, $\bm{a}$ are linear factors to be extracted, connected via a hierarchical model with no error in the second stage of hierarchy. Linear ICA can be useful here as it separates the observed data  into a lower-dimensional array of independent sources (a.k.a. features) to offer a high-dimensional probabilistic structure. Thus, ICA falls into the class of high-dimensional data reduction methods and help one identify pivotal data features as well as yield optimal predictive outcomes. A key feature of popular statistical or machine learning models is that they necessitate an approach that concurrently discerns both the foundational independent features and their associated mixing weights, colloquially termed as nonlinear mixing maps.

Our primary objective is to unify an array of algorithmic approaches to allow for both MAP optimization and fully Bayesian estimation for ICA.  Method of moments estimators are commonplace in this literature and have lead to tensor methods.  Our work builds on the seminal work of \citep{mackay1992bayesian, mackay1996maximum} who showed that the \citep{bell1995information} algorithm for signal processing for a non-linear feedforward network can be viewed as a maximum likelihood algorithm for the optimization of a linear generative model. 

Our main contributions are: 
\begin{enumerate*}
\item Developing a fully Bayesian ICA framework with horseshoe-type priors, using a \PG{}-augmented Gaussian scale mixture hierarchy that leads to a simple Gibbs sampler (\texttt{Gibbs-ICE}) for posterior inference on the sources and the unmixing matrix.
\item Unifying existing models using Gaussian scale mixture representation. 
\item Providing, to our knowledge, the first posterior contraction and Bernstein--von Mises results for hierarchical Bayesian ICA: under known source densities and mild regularity conditions, we establish a uniform LAN expansion for the ICA likelihood and show that the posterior for the unmixing matrix $W$ contracts at the parametric rate $N^{-1/2}$ up to signed permutations, and is asymptotically normal (Section~\ref{sec:theory}).
\item Carrying out an extensive simulation study, including controlled experiments under our generative hierarchy and large-scale benchmarks against contemporary ICA methods. Using a Gibbs sampler (\texttt{Gibbs-ICE}) built from our hierarchy, we find that the proposed Bayesian procedures are competitive in terms of Amari distance, source recovery correlation, and reconstruction error across a range of super-Gaussian source families (Sections~\ref{sec:app}--\ref{sec:contemporarymethods}).
\end{enumerate*}

The hallmark of ICA methods is the use of a super-Gaussian distribution over hidden states. Surprisingly, this leads to identification where traditional Gaussian modeling assumptions do not.  A linear mixing of Gaussian distribution is itself Gaussian, so de-mixing is impossible. Hence the source distribution must have heavy tails (a.k.a. super-Gaussian).
We show that the easiest way to achieve this is via `Horseshoe-type' priors which are heavy-tailed by construction. Horseshoe priors have been a default class of priors for many problems, especially shrinkage estimation in high-dimensional inference, where the marginal density possesses both a spike at zero and regularly varying tails, see \citet{carvalho2010horseshoe, bhadra_default_2016, bhadra2017horseshoe}. 

Generative models have achieved many successful applications ranging from machine learning to image processing. They can be applied in the context of ICA. To fix notation let the data generating model be constructed as follows: data $\bm{y}$ is generated from a baseline source distribution, $\bm{s} \sim p(\bm{s})$ via a latent state $\bm{x}$ which is a ridge function of $\bm{s}$, namely $ g(\bm{A}\bm{s} + \b)$ for an activation function $g(\cdot)$ and then, 
\[
\bm{x} = g(\bm{A} \bm{s}  + \bm{b}), \;  {\rm where} \; \;  \bm{s} \sim p(\bm{s}). 
\]
The mixing matrix $\bm{A}$ provides `feature extraction' -- a central problem in machine learning. We consider below the special case $g = \mathbbm{1}(\cdot)$, the identity map, but $g$ could be a deep learner with many layers. Training high-dimensional generative models such as LLMs (large language models) is extremely costly and novel scalable algorithms to help in this task are an active area of interest. We refer the readers to \citet{polson2020deep, tran2020bayesian} for a comprehensive review of the computational aspects of deep learning. 

Our approach is inspired by the seminal insight of \citet{mackay1992bayesian, mackay1996maximum} that ICA can be viewed as a latent variable modeling problem. \citet{mackay1996maximum} provides instances of many statistical models (e.g., mixture models, hidden Markov models, factor analysis or Helmholtz machines) that are generative models with a layer of latent variables that are usually modeled with simple, separable distributions. Across each of these modeling frameworks, learning the latent variables is tantamount to describing the observables in terms of independent components, and hence, it is natural to expect that ICA should also admit a generative latent variable representation. With this interpretation, \citet{mackay1992bayesian} shows that the original blind separation algorithm of \citet{bell1995information} can be viewed as a maximum a posteriori (MAP) estimator from the marginalized likelihood, where the trick is to write the ICA as a latent variable model with a separable distribution on the hidden states. In this representation, the latent variables are assumed to be mutually independent and non-Gaussian, and are called independent components. The key insight then that underlies our unifying approach is that the components of the source  distribution can be modeled with scale mixtures of normals \citep{west1987scale, barndorff1982normal}.  We show that MacKay's original model and its extensions are simply scale mixtures with P\'olya-Gamma mixing \citep{polson2013bayesian} and are thus horseshoe-type priors \citep{polson2012half,polson_data_2013}.

On the algorithmic side, we show that a number of hitherto disjoint algorithms can be unified as envelope optimization methods \citep[see][]{geman1995nonlinear, polson2015mixtures}. For example, the ICA updating scheme due to \citet{bell1995information} is equivalent to a maximum likelihood approach, see \citep{mackay1992bayesian, mackay1996maximum}. Auxiliary variable methods allow for both EM and MCMC algorithms to be developed across a wide spectrum of source distributions. Such methods can lead to faster convergence \citep{ono2010auxiliary} and can incorporate methods for fast convergence such as Nesterov acceleration and block-coordinate descent, thus providing an alternative to traditional stochastic gradient descent methods. 

From a historical perspective,  Independent Component Analysis arises as a possible solution to the `Blind source separation' problem, a classical problem in signal processing. In the formulation by \citep{herault1986space, bell1995information}, algorithms for blind source separation attempt to recover source signals $\bm{s}$ from observations $\bm{x}$, where $\bm{x} = \bm{V} \bm{s}$, are linear mixtures with unknown weights $\bm{V}$. This is done by finding a square matrix $\bm{W}$ which is the inverse of the mixing matrix $\bm{V}$, up to permutation and change of scale. For example, \citep{bell1994non, bell1995information} take an algorithmic approach summarized as a linear mapping $\bm{a} = \bm{W} \bm{x} $ where $\bm{a}$ is an estimate of the source signals. This is done by adjusting the unmixing matrix $\bm{W}$ to maximize the entropy of the outputs. 

The algorithm proceeds iteratively using a gradient ascent method on the log-likelihood of the estimated sources, where the update rule for $\bm{W}$ is:
\[
\nabla \bm{W} \propto \left( I - 2\phi(\bm{a}) \bm{a}^T \right) \bm{W}
\]
where, $I$ is the identity matrix, and $\phi(\bm{a})$ is a non-linear function applied component-wise to $\bm{a}$. Common choices for $f$ are the logistic sigmoid function or the hyperbolic tangent (i.e., a nonlinear map $\z_i = \phi_i (\bm{a}_i ) $ where $ \phi(\cdot) = - \tanh(\cdot)$. 

A Bayesian approach \citep{fevotte2004bayesian, fevotte2006blind, donnat2019constrained, mackay1992bayesian} has a two-fold advantage. First, rather than casting problem as a method of moments approach (kurtosis), a Bayesian approach finds the solution by a suitable regularization with  a heavy-tailed prior (thus imposing constraint on higher order moments). A particular suitable class is super-Gaussian priors \citep{palmer2006super} via scale mixtures of normals \citep{west1987scale}. \citet{mackay1992bayesian} uses a heavy-tailed source distribution towards this. The advantage is that mixture and envelope methods can be used to extract features/factors and provide fast scalable algorithms for ICA. In this paper, we unify the existing Bayesian algorithms and provide new ones.

Assuming a  source distribution,  $\bm{s} \sim p(\bm{s}) $ acts as a regularization penalty and allows us to unify existing procedures as MAP estimators.  For example,  \citet{mackay1996maximum} shows that if $\phi_i(a_i) = - \tanh (a_i)$ then this is equivalent to a source distribution of the form $ p_i(s_i) \propto 1 / \cosh(s_i) = 1/(\e^{s_i}+\e^{-s_i})$. We argue in section \ref{sec:model} that this can be written as a Gaussian scale mixture using a \PG{} mixing density. Moreover, \citep{mackay1992information} suggests adding a gain $\beta $ and considering a heavy-tailed source distribution of the form $p_i(s_i ) \propto 1 / \cosh^{1/\beta}( \beta s_i )$. Again we show that this can be represented as a Normal scale mixture thus leading to classes of auxiliary variable methods for inference and optimisation. In the limit as $\beta \to \infty $ this becomes $ p_i(s_i) \propto \exp (-\abs{s_i})$ the double exponential (a.k.a. Lasso) prior, and as $\beta \to 0$, $p_i(s_i)$ would converge to a zero-mean Gaussian with variance $1/\beta$. 


A related goal is to bring together the modeling of source distributions under the encompassing paradigm of Gaussian scale mixtures \citep{andrews1974scale, west1987scale, carlin1991inference}. This latent variable or parameter expansion approach offers fast, scalable algorithms akin to the auxiliary variable methods proposed by \citep{ono2010auxiliary}. Furthermore, it obviates dependence on approximative techniques like variational Bayes (VB). Our methodology bears significance for deep Bayesian models wherein latent variable distributions are forged as superpositions of nonlinear mappings, akin to deep learning constructs \citep{polson2017deep, polson2018posterior}.

On the theoretical side, Section~\ref{sec:theory} makes this link precise by formulating the ICA likelihood as a finite-dimensional parametric model indexed by the unmixing matrix and proving a uniform LAN expansion around the true $W_0$. This leads to $\sqrt{N}$-rate posterior contraction in a signed-permutation invariant metric $d_\pm$ and a Bernstein--von Mises theorem for $\mathrm{vec}(W)$, thus providing a matrix-level asymptotic normality result for ICA in a hierarchical Bayesian framework. On the empirical side, Section~\ref{sec:app} and Section~\ref{sec:contemporarymethods} show that our \PG-augmented Gibbs sampler gives practically competitive performance relative to established ICA algorithms across a range of super-Gaussian source distributions and noise levels.

The rest of our paper is outlined as follows. The next subsection describes connections with previous work. Section \ref{sec:model} provides a discussion of deep Bayesian generative models. Section \ref{sec:super} provides a unifying framework for super-Gaussian source distributions. Section \ref{sec:theory} develops our posterior contraction and Bernstein--von Mises results for the unmixing matrix. Section \ref{sec:app}--\ref{sec:contemporarymethods} illustrate our methodology through numerical experiments and extensive comparisons with contemporary ICA methods. Finally, Section \ref{sec:discuss} concludes with directions for future research. 

\subsection{Connections with Previous Work}

As \citet{auddy2023large} point out, the mixing matrix $\bm{A}$ can be identified up to permutation and scaling under the assumption of at most one Gaussian component in $\S$ due to a theorem by \citet{comon1994independent}. The review paper \citet{auddy2024tensor} demonstrates that this can also be connected with the Kruskal identifiability theorem \citep{kruskal1977three} that shows a remarkable difference between matrix and tensors: while a matrix of rank $\rho$ can be decomposed in many ways as a sum of $\rho$ rank-one matrices, but higher-order tensors typically admit unique decomposition into rank-one tensors. In the context of PCA versus ICA, \citet{auddy2024tensor} observe that the the principal components are singular vectors of covariance matrix, while the independent components are `singular vector's of fourth order cumulant tensors, and hence, PCA can not reconstruct independent random variables with mean zero, unit variance but non-zero excess kurtosis, but ICA can. There have been many approaches for solving the ICA problem including but not limited to \citep{mackay1992bayesian, comon1994independent, bell1995information, mackay1996maximum, deco1996information, hyvarinen1999nonlinear, hastie2002independent, ono2010auxiliary, samworth2012independent,auddy2023large} as well as hierarchical Bayesian approaches such as \citet{mackay1992bayesian, fevotte2004bayesian, fevotte2006blind, karklin2005hierarchical, asaba2018bayesian}. We refer the readers to \citet{hyvarinen2001independent, nordhausen2018independent} for comprehensive reviews. 

\citet{samworth2012independent} proposes a nonparametric maximum likelihood approach for estimating the unmixing weights and the marginal distributions by projecting the empirical distributions on the space of log-concave univariate distributions. They also address the question of identifiability of the unmixing matrices, first studied by \citet{comon1994independent}, and in their formulation identifiability up to scaling and permutation is guaranteed by requiring that not more than one of the univariate log-concave projections is Gaussian. This is also connected to Kruskal's identifiability conditions \citep{kruskal1977three} for a tensor product $T = \sum_{r =1}^{R} A \otimes B \otimes C$, in terms of the Kruskal ranks of the individual matrices: $\kappa_A + \kappa_B + \kappa_C \ge 2R +2$. \citet{bhaskara2014uniqueness} provide a robust version of the identifiability theorem for approximate recovery and point uses in several latent variable models. We refer the readers to the comprehensive review by \citet{auddy2024tensor} on tensor methods for high-dimensional data analysis. 

\citet{camuto2021towards} combine a linearly independent component analysis (ICA) with nonlinear bisective feature maps (from flow-based methods). For non-square ICA, they can assume the number of sources is less than the data dimensionality -- thus achieving better unsupervised latent factor discovering than other ICA flow-based methods.  \citet{dinh2014nice} discuss nonlinear independent component estimation (NICE), deep generative models with nonlinear invertible neural networks, building on the work of \citet{deco1996information}, \citet{obradovic1998information}, \citet{comon1991blind} and \citet{pearlmutter1996context} and \citet{malthouse1998limitations}. Recent work includes Bayesian ICA models of \citep{donnat2019constrained}. Other related approaches include auto-encoder and sparsity models. 

Nonlinear ICA models have been proposed by \citet{hyvarinen1999nonlinear}, although identification can be challenging. \citet{khemakhem2020variational} provides recent identification results, providing mild conditions under which the joint distribution encompassing both observed and latent variables within Variational Autoencoders (VAEs) are identifiable and estimable, thus establishing a connection between VAEs and nonlinear Independent Component Analysis (ICA).
The source $\S$ is a feature vector that needs to be learned, see, for example, \citet{olshausen1996emergence, olshausen1997sparse} for sparse coding of natural images, where an image is modeled as a natural superposition using an over-complete basis set where the amplitudes are given sparsity-inducing prior distributions. This is based on the intuition of Barlow's principle of redundancy reduction \citep{barlow1989unsupervised, barlow2001redundancy}. 

Other popular approaches for dimension reduction include Sliced Inverse Regression \citep{li1991sliced} that finds a low-dimensional projection of the data that captures the most relevant information for explaining the variation in the data, specifically designed for non-linear relationships in the data, Unlike traditional linear dimensionality reduction methods like Principal Component Analysis (PCA). \citet{lopes2012bayesian} introduces a sequential online strategy for efficient posterior simulation. Finally,  as noted by \cite{brillinger_generalized_2012} and \cite{Naik2000}, the mixing matrix can be consistently estimated through PLS, regardless of the activation function's nonlinearity, albeit with a proportionality constant. While the assumption by \cite{brillinger_generalized_2012} of Gaussian input $\bm{X}$ is necessary for applying Stein's lemma, we note that this outcome extends to scale-mixtures of Gaussians.

\section{Hierarchical Independent Components Analysis}\label{sec:model}
\subsection{The generative model and the likelihood function}

We begin with linear independent component models. The goal of ICA is to attempt to recover source signals $\bm S$ from observations $\bm X$ which are linear mixtures (with unknown coefficients $\bm A$) of the source signals, i.e.\ $\bm X = \bm A \bm S$, where $\bm W = \bm A^{-1}$. This can be interpreted as a Bayesian hierarchical model with a degenerate first stage. We follow \citet{mackay1996maximum}'s formulation here. We observe data as $N$ observations $\bm x^{(n)}$, $1\leq n \leq N$, which are linear mixtures of sources $\bm s^{(n)}$. Let $\bm s^{(n)} \in \mathbb{R}^d$, $1\leq n \leq N$, denote the set of sources, that are independently distributed with marginal density $p_i(s_i^{(n)})$, so that
\[
\bm x^{(n)} = \bm A \bm s^{(n)}  
\quad \text{with} \quad  
p(\bm s^{(n)}) = \prod_{i=1}^d p_i(s_i^{(n)}),
\]
where $\bm A$ is the mixing matrix. We wish to estimate $\bm W=\bm A^{-1}$. As stated before, the sources $\bm s^{(n)}$ have an independent components distribution and can be viewed as latent variables. The joint probability of the observed and the hidden latent variables can be written as 
\begin{align}
p\big(\{\bm x^{(n)}\}_{n=1}^{N}, \{\bm s^{(n)} \}_{n=1}^{N} \mid \bm A\big) 
&= \prod_{n=1}^{N} p(\bm x^{(n)} \mid \bm s^{(n)}, \bm A)\, p(\bm s^{(n)}) \nonumber \\
&=  \prod_{n=1}^{N} 
\Bigg\{\prod_{i=1}^d \delta\Big(x^{(n)}_i - \sum_{j=1}^d A_{ij}s^{(n)}_j\Big)\Bigg\}
\prod_{j=1}^d p_j(s_j^{(n)}), \label{eq:joint}
\end{align}
where $\delta(\cdot)$ denotes the Dirac delta function.
\citet{mackay1996maximum} points out that it is straightforward to replace $\delta( x^{(n)}_j - \sum_i A_{ji} s_i^{(n)})$ with a probability distribution over $x_j^{(n)}$ with mean $\sum_i A_{ji} s_i^{(n)}$, but we need to assume $\bm x$ is generated without noise to obtain the \citet{bell1995information} algorithm exactly.

To access a wider range of probability densities, we can introduce auxiliary variables $\lambda$ such that
\[
p(\bm s^{(n)}) = \int p(\bm s^{(n)} \mid \lambda)\,p(\lambda)\, d\lambda. 
\]
Gaussian scale mixtures \citep{west1987scale, bhadra2016global} encompass a wide range of commonly used prior distributions in Bayesian literature, and are also a source of constructing newer priors for handling data with sparsity or other structures. In fact, it is easy to see that the existing prior distributions for Bayesian ICA, viz.\ Student's $t$ \citep{fevotte2004bayesian}, Jeffreys' prior $p(s_i) \propto 1/\abs{s_i}$ \citep{fevotte2006blind}, or Laplace \citep{asaba2018bayesian}, are all Gaussian scale mixtures. We will argue later in the section that one can recover \citet{mackay1992bayesian}'s hyperbolic secant distribution as a Gaussian scale mixture, too. This has many theoretical and practical advantages. For example, in designing computational algorithms, we can develop EM and MCMC algorithms using the joint posterior $p(\bm s^{(1:N)} , \lambda \mid \bm x^{(1:N)})$. We also know a great deal about the behavior of Gaussian scale mixtures due to the works of \citet{barndorff1982normal} and others. 

To obtain the maximum likelihood estimator under the noiseless model, we first observe that the likelihood can be obtained as a product of the following factors from \eqref{eq:joint}, for $n = 1, \ldots, N$:
\begin{gather*}
p( \bm x^{(n)} \mid  \bm A )  
= \int p(\bm x^{(n)} \mid \bm A ,\bm s^{(n)})\, p(\bm s^{(n)})\, d \bm s^{(n)} 
= \int \Bigg\{\prod_{i=1}^d \delta\big( x^{(n)}_i - \sum_{j=1}^d A_{ij} s^{(n)}_j \big)\Bigg\}
\prod_{i=1}^d p_i(s_i^{(n)})\, d \bm s^{(n)}.
\end{gather*}
We use the summation convention like \citet{mackay1996maximum}, i.e., $A_{ij} s_j^{(n)} \equiv \sum_{j=1}^d A_{ij} s_j^{(n)}$. Now, using the elementary fact that 
\[
\int \delta(\bm x - \bm A \bm s)\, f(\bm s)\, d\bm s = |\bm A|^{-1} f(\bm W \bm x),
\quad \bm W = \bm A^{-1},
\]
we obtain the likelihood (and log-likelihood) for a single term:
\begin{gather*}
p( \bm x^{(n)} \mid \bm A ) 
= \frac{1}{|\bm A|} \prod_{i=1}^d  p_i \big((\bm A^{-1}\bm x^{(n)})_i\big)
=  \prod_{i=1}^d  p_i\big( W_{ij} x^{(n)}_j \big)\, |\bm W|, \\
\Rightarrow \log p(\bm x^{(n)} \mid \bm A) 
= \log |\bm W| + \sum_{i=1}^d \log p_i\big(W_{ij} x^{(n)}_j\big) 
\doteq \log |\bm W| + \sum_{i=1}^d \log p_{i}(a_i), 
\ \  \text{where } a_i \equiv W_{ij} x^{(n)}_j. 
\end{gather*}
We can then optimize the log-likelihood using any gradient or envelope method. The terms $\phi_i(a_i) = d \log p_i(a_i) / d a_i$ are key here and indicate the gradient direction for maximum likelihood. 

We mention two key points in \citet[(Section 2.4, points 2 and 3)]{mackay1996maximum} here:
\begin{enumerate}
    \item Employing a $\tanh$ nonlinearity of the form $\phi_i(a_i) = -\tanh(a_i)$ implicitly assumes a probability distribution for latent variables, $p_i(s_i) \propto {1}/{\cosh(s_i)} \propto {1}/({e^{s_i} + e^{-s_i}})$. This distribution exhibits heavier tails compared to the Gaussian distribution, offering a broader range of potential tail behaviors.

    \item Alternatively, by incorporating a $\tanh$ non-linearity with a gain parameter $\beta$, denoted as $\phi_i(a_i) = -\tanh(\beta a_i)$, the associated probabilistic model varies with $\beta$. The resulting distribution is expressed as $p_i(s_i) \propto {1}/{[\cosh(\beta s_i)]^{1/\beta}}$. As $\beta$ becomes large, the non-linearity converges to a step function, resulting in a Laplace density $p_i(s_i) \propto \exp(-|s_i|)$. Conversely, as $\beta$ approaches zero, $p_i(s_i)$ tends towards a Gaussian distribution with a mean of zero and a variance of $\frac{1}{\beta}$.
\end{enumerate}

\begin{remark}{\textbf{Time versus Transfer Domain}.}
It is worthwhile to note that separation in transfer domain is equivalent to separation in the time domain due to the one-to-one mapping between the models in time and transfer domain. Using the notations in \citet{fevotte2006blind, fevotte2004bayesian}, the linear instantaneous model in time domain is given by the following model where observations at time $t$ are noisy combinations of sources at $t$, for signals of length $N$:
\beq
\bm x_t = \bm A \bm s_t + \bm\varepsilon_t, \; t = 0, \ldots, N-1, \label{eq:timedomain}
\eeq
where $\bm x_t = (x_{1t}, x_{2t}, \ldots, x_{dt})^\top$ are the observation vectors, $\bm s_t = (s_{1t}, s_{2t}, \ldots, s_{dt})^\top$ are the sources and $\bm\varepsilon_t = (\epsilon_{1t}, \epsilon_{2t}, \ldots, \epsilon_{dt})^\top$ are the additive noises. The goal is to estimate $\bm s_t$ and $\bm A$. We assume that there is a basis $N \times N$ matrix $\Phi$ such that sources have a sparse representation on it. The equivalent model in the transfer domain is then
\beq
     \bm x^{(n)} = \bm A \bm s^{(n)} + \bm\varepsilon^{(n)}, \; n = 0, \ldots, N-1, \label{eq:transferdomain}
\eeq
where $n$ is the coefficient index in the basis decomposition. Separation using \eqref{eq:timedomain} and \eqref{eq:transferdomain} are equivalent since $\Phi$ is a basis matrix. 
\end{remark}

\subsection{Exponential family representation and the posterior} 

We now show the general derivation for the posterior under the aforementioned hierarchical model for the observables and the latent variables in a noisy linear setting. To reduce notational clutter, we drop the superscript $(n)$ and adopt summation convention as needed. Consider
\[
\bm x = \bm A \bm s + \bm\varepsilon,\quad \bm\varepsilon \sim \mathcal{N}_d(\bm 0,\sigma^2 \bm I_d),
\]
with prior $p(\bm s)$ for the sources. The likelihood $p(\bm x \mid \bm s)$ derived from this model together with the source distribution $p(\bm s)$ can be combined to form the posterior of $\bm s$ given $\bm x$:
\begin{align*}
p(\bm s \mid \bm x) 
&= \frac{1}{(2\pi \sigma^2)^{d/2}}
\exp\left\{-\frac{1}{2\sigma^2}(\bm{x - A s})^\top(\bm{x - A s})\right\} p(\bm s)\\
&= h(\bm s)\, \exp\left\{\bm{\eta}^\top \bm s - \frac{1}{2\sigma^2}\bm{x}^\top\bm{x} -\log f(\bm x) \right\}\\
&= h(\bm s)\, \exp\left\{\bm{\eta}^\top \bm s - \psi(\bm\eta) \right\},
\end{align*}
where the different components (known functions) are given by
\begin{align*}
\bm\eta & = \frac{1}{\sigma^2}\bm{A}^\top \bm x 
\quad\quad (\text{canonical parameter}), \\
\psi(\bm \eta) &= \frac{1}{2\sigma^2}\bm{x}^\top\bm{x} +\log f(\bm x) 
= \frac{\sigma^2}{2}\bm{\eta}^\top\bm{W W}^\top\bm{\eta} + \log f(\sigma^2 \bm{W}^\top\bm{\eta})
\quad (\text{cumulant function}), \\ 
\text{where}\quad 
f(\bm x) &= \int  h(\bm s)\, 
\exp\left\{\bm{\eta}^\top\bm s - \frac{1}{2\sigma^2}\bm{x}^\top\bm{x} \right\} d\bm s, \quad \text{and}\\
h(\bm s) &= \frac{1}{(2\pi \sigma^2)^{d/2}} 
\exp\left\{-\frac{1}{2\sigma^2} \bm{s}^\top\bm{A}^\top\bm{A}\bm{s}\right\} p(\bm s). 
\end{align*}
Therefore, we can calculate the posterior mean of $\bm s$ given $\bm x$ as
\begin{gather*}
\E(\bm s \mid \bm x) 
= \frac{\partial\psi(\bm\eta)}{\partial \bm\eta} 
= \bm{W}\bm x + \sigma^2 \bm W \frac{\partial}{\partial \bm x}\log f(\bm x) 
= \frac{\displaystyle\int h(\bm s)\exp\{\bm{\eta}^\top\bm s\} \bm s\, d\bm s}
{\displaystyle\int h(\bm s)\exp\{\bm{\eta}^\top\bm s\} d\bm s}, \\
\frac{\partial}{\partial \bm x}\log f(\bm x) 
= \frac{1}{\sigma^2}\left(\bm{A} \frac{\displaystyle\int h(\bm s)\exp\{\bm{\eta}^\top\bm s\} \bm s\, d\bm s}
{\displaystyle\int h(\bm s)\exp\{\bm{\eta}^\top\bm s\} d\bm s} - \bm x\right). 
\end{gather*}
We will use this later when deriving a full MCMC algorithm in Appendix \ref{sec:env} for uncertainty in ICA models.

\section{Super-Gaussian Source Distributions}\label{sec:super}
As discussed earlier, it is well known that one needs to use a heavy-tailed super-Gaussian distribution as prior for the sources $\bm s$ to identify the mixing matrix $\bm A$. In addition, as \citet{fevotte2004bayesian} point out, while the over-determined case (number of sensors $\ge$ number of sources) is relatively easy: there are many efficient approaches, especially within Independent Component Analysis; the general linear instantaneous case, with mixtures possibly noisy and under-determined (number of sensors $\le$ number of sources) is challenging. A common approach in such situations is sparsity-based Blind Source Separation (BSS), especially for under-determined mixtures. The basic premise is to exploit source sparsity: only a few expansion coefficients of sources are significantly different from zero. For example, recommendations in the literature include Student's $t$ or Jeffreys' prior on $\bm s$ \citep{fevotte2004bayesian, fevotte2006blind} or Laplace \citep{asaba2018bayesian}, all of which are (or can be written as) Gaussian scale mixtures. We take a Bayesian shrinkage approach: the prior $\bm s \sim p(\bm s)$ acts as a regularization penalty and allows us to unify existing procedures as MAP estimators.

\subsection{MacKay source distribution: hyperbolic secant}

\citet{mackay1996maximum} shows that the choice $\phi_i(a_i) = - \tanh(a_i)$ is equivalent to a source distribution of the form
\[
p_i(s_i) \;\propto\; \frac{1}{\cosh(s_i)} \;=\; \frac{1}{\e^{s_i}+\e^{-s_i}},
\]
the (unstandardized) hyperbolic secant distribution. This can be written as a normal scale mixture using a \PG{} mixing density:
\begin{equation}
  s \mid \tau \sim \NormRV\!\left(0, \frac{1}{4\tau}\right),\quad 
  \tau \sim PG(1,0),
  \label{eq:sech-mixture}
\end{equation}
since there exists a density $p(\tau)$ on $\mathbb{R}^+$ such that
\begin{equation}
  \frac{1}{\cosh(s)} \;\propto\; \int_{0}^{\infty} \exp\!\big(-2\tau s^2\big)\,p(\tau)\,d\tau.
\end{equation}
Therefore, the location parameter $s$ is a normal scale mixture, and we can further introduce a global scale $v$ if desired:
\begin{align}
s \mid v,\tau \sim \NormRV\!\left(0,\ \frac{v^2}{4\tau}\right),
\quad v\sim p(v),\quad \tau\sim PG(1,0).
\end{align}

To see the connection to the \PG{} distribution more explicitly, recall that:

\begin{definition}
  Suppose $\beta \ge 0$. The \PG{} distribution $PG(\beta)$ is defined as the density $p_{PG}$ on $\mathbb{R}^+$ whose Laplace transform satisfies
  \[
    \cosh^{-\beta}\!\left(\sqrt{\frac{t}{2}}\right) \;=\;
    \int_0^{\infty} \exp(-tx)\, p_{PG}(x \mid \beta)\,dx .
  \]
\end{definition}

Setting $\beta=1$ and choosing $t$ appropriately yields the scale mixture representation for $1/\cosh(s)$. This is also connected to Jacobi theta distributions studied by \citet{biane2001probability, devroye2009exact}: $J^*$ has a Jacobi distribution if
\[
  J^* \;\stackrel{d}{=}\; \frac{2}{\pi^2} \sum_{k = 1}^{\infty} \frac{e_k}{(k-\tfrac12)^2},
  \qquad e_k \sim \ExpRV(1),
\]
with moment generating function
\[
  \E\big(\e^{-tJ^*}\big) = \frac{1}{\cosh\sqrt{2t}}.
\]

The $\cosh(\cdot)$ priors also have a connection with the popular horseshoe priors:
\[
  \beta_j \sim \NormRV(0, \lambda_j^2 \tau^2 \sigma^2),
  \quad \lambda_j \sim p(\lambda_j)\, d\lambda_j,
  \quad \tau \sim p(\tau)\, d\tau.
\]
The horseshoe prior places a standard half-Cauchy distribution over each local scale $\lambda_j$. This induces an unstandardized unit hyperbolic secant distribution over $\xi_j = \log(\lambda_j)$ \citep[see][Section~5]{polson_halfcauchy_2012}:
\[
  p_{HS}(\xi_j) = \frac{1}{\pi}\,\frac{1}{\cosh(\xi_j)}.
\]
Such log-scale shrinkage priors can unify many commonly used continuous shrinkage priors, as shown in \citet{schmidt2018log}. This also provides a link between the non-Gaussian tail behavior needed for blind source separation and that of sparse signal extraction. A similar framework to horseshoe and optimal rates of reconstruction is an area of future research.

Other proposals include the scaled $\cosh$ prior \citep{mackay1996maximum}: $p(s) \propto 1/\cosh^{1/\beta}(\beta s)$, which has the interesting limiting behavior
\[
    p_i(s_i \mid \beta) \;\propto\;
    \begin{cases}
      \exp(-\abs{s_i}) & \beta \to \infty \quad \text{(Bayesian Lasso)} ,\\[2pt]
      \NormRV(0, 1/\beta) & \beta \to 0 \quad \text{(Bayesian Ridge)} .
    \end{cases}
\]
That is, in the limit $\beta \to 0$ or $\beta \to \infty$, we recover the Lasso or its Bayesian analog \citep{tibshirani96, park_bayesian_2008}, as noted by \citet{mackay1996maximum}.

\subsection{Gibbs sampling strategies}

Our main computational contribution is to exploit the \PG{}-augmented representation in \eqref{eq:sech-mixture} to derive a simple, fully conjugate Gibbs sampler for ICA under the hyperbolic secant source prior.

For concreteness, consider a noisy linear ICA model with $N$ observations and $d$ components:
\[
  \bm x_i \mid \bm s_i, \bm A \;\sim\; \NormRV_d(\bm A \bm s_i,\ \sigma^2 \bm I_d),
  \qquad i=1,\dots,N,
\]
where $\bm A \in \mathbb{R}^{d\times d}$ is the mixing matrix and $\bm s_i \in \mathbb{R}^d$ are the sources. Under the MacKay hyperbolic secant prior with \PG{} augmentation,
\[
  s_{ij} \mid \tau_{ij} \sim \NormRV \left(0,\ \frac{1}{4\tau_{ij}}\right),
  \quad \tau_{ij} \sim PG(1,0),
\]
for $i=1,\dots,N$, $j=1,\dots,d$.
For the mixing matrix, we may place independent Gaussian priors on the columns,
\[
  \bm a_k \sim \NormRV_d(\bm 0,\ \sigma_2^2 \bm I_d), \quad
  k=1,\dots,d,
\]
where $\bm a_k$ is the $k$-th column of $\bm A$ and $\sigma_2^2>0$ is a prior variance.

Let $\bm X\in\mathbb{R}^{N\times d}$ collect the observations row-wise, $\bm S\in\mathbb{R}^{N\times d}$ collect the sources, and $\bm T\in\mathbb{R}^{N\times d}$ collect the latent \PG{} scales $\tau_{ij}$. The joint posterior $p(\bm S,\bm A,\bm T \mid \bm X)$ is then amenable to a three-block Gibbs sampler, cycling through:

\paragraph{1. Update the sources $\bm S \mid \bm A,\bm T,\bm X$.}
Conditionally on $(\bm A,\bm T)$, the rows of $\bm S$ are independent. For each $i=1,\dots,N$ let
\[
  \bm D_i = \mathrm{diag}(4\tau_{i1},\dots,4\tau_{id}),\qquad
  \bm\Sigma_{s,i} = \Big(\sigma^{-2}\bm A^\top \bm A + \bm D_i\Big)^{-1},
\]
and
\[
  \bm\mu_{s,i} = \sigma^{-2} \bm\Sigma_{s,i}\,\bm A^\top \bm x_i.
\]
Then
\[
  \bm s_i \mid \bm A,\bm T,\bm X \;\sim\; \NormRV_d\big(\bm\mu_{s,i},\ \bm\Sigma_{s,i}\big).
\]

\paragraph{2. Update the mixing matrix $\bm A \mid \bm S,\bm X$.}
Given $\bm S$, the columns of $\bm A$ are conditionally independent. For $k=1,\dots,d$, consider the $k$-th column of $\bm X$,
$\bm x_{\cdot k} \in \mathbb{R}^N$, and write
\[
  \bm x_{\cdot k} = \bm S \bm a_k + \bm\varepsilon_{\cdot k},\qquad
  \bm\varepsilon_{\cdot k} \sim \NormRV_N(\bm 0,\ \sigma^2 \bm I_N).
\]
Combining this likelihood with the Gaussian prior yields the full conditional
\[
  \bm\Sigma_{a} = \Big(\sigma^{-2}\bm S^\top \bm S + \sigma_2^{-2}\bm I_d\Big)^{-1},
  \qquad
  \bm\mu_{a,k} = \sigma^{-2}\bm\Sigma_{a}\,\bm S^\top \bm x_{\cdot k},
\]
so that
\[
  \bm a_k \mid \bm S,\bm X \;\sim\; \NormRV_d\big(\bm\mu_{a,k},\ \bm\Sigma_{a}\big),
  \qquad k=1,\dots,d.
\]

\paragraph{3. Update the latent \PG{} scales $\bm T \mid \bm S$.}
Conditionally on $\bm S$, the entries $\tau_{ij}$ are independent and follow \PG{} full conditionals:
\[
  \tau_{ij} \mid \bm S,\bm X,\bm A \;\sim\; PG\big(1,\ \abs{2s_{ij}}\big),
  \qquad i=1,\dots,N,\ j=1,\dots,d.
\]

Putting these steps together, the Gibbs sampler iteratively updates
\[
  (\bm S^{(m)},\bm A^{(m)},\bm T^{(m)})
  \quad\longrightarrow\quad
  (\bm S^{(m+1)},\bm A^{(m+1)},\bm T^{(m+1)})
\]
by drawing successively from the three full conditionals above.
This \PG{}-augmented Gibbs scheme is the core of our proposed \texttt{Gibbs-ICE} algorithm and is our main computational contribution: it yields a simple, fully conjugate MCMC method for ICA under a principled hyperbolic secant or horseshoe-type prior, while retaining exact Bayesian updating. The algorithmic steps are outlined in Algorithm \ref{alg:gibbs-ice}.

\paragraph{Jeffreys prior.}
For completeness, we briefly recall the Jeffreys prior used in \citet{fevotte2006blind}. This corresponds to the improper source distribution $p(s) \propto 1/\abs{s}$. \citet{fevotte2006blind} derive an EM algorithm for MAP estimation under Jeffreys' prior using the following Normal scale mixture representation:
\[
  \frac{1}{\abs{s_{ik}}} = \int \NormRV(s_{ik} \mid 0, v_{ik})\, \frac{1}{v_{ik}}\, dv_{ik}.
\]
The prior $p(v_{ik}) \propto 1/v_{ik}$ can be viewed as a limiting case of the inverse-gamma prior as both parameters approach zero, since
$p(v) \propto v^{-(a+1)} \e^{b/v} \to 1/v$ as $a,b \to 0$. It then follows that the prior $p(s_{ik}) \propto 1/\abs{s_{ik}}$ can be derived as a limiting case of the $t$-prior in \citet{fevotte2004bayesian}.

\paragraph{Student's $t$ prior \citep{fevotte2004bayesian}.}
In the presence of noise, $\bm x_j = \bm A \bm s_j + \bm\varepsilon_j$, $j = 1,\ldots,N$, we still obtain a Gaussian scale mixture framework and EM or Gibbs algorithms can be constructed. For example, \citet{fevotte2004bayesian} provide MCMC algorithms by iterating the conditionals where each $s_i$ has a $t(\alpha_i,\lambda_i)$ prior with degrees of freedom $\alpha_i$ and scale $\lambda_i$, which can be written as a Normal scale mixture of inverse-gamma distributions:
\[
  p(s_{ij} \mid v_{ij}) = \NormRV(s_{ij} \mid 0, v_{ij}),
  \quad
  p(v_{ij} \mid \alpha_i, \lambda_i) = \InvGammaRV\Big(v_{ij} \,\Big|\, \frac{\alpha_i}{2},\ \frac{2}{\alpha_i \lambda_i^2}\Big).
\]
The resulting Gibbs sampler iterates over:
\begin{itemize}
  \item Source update:
  letting $\bm\Sigma_{s_j} = \big(\sigma^{-2} \bm A^\top \bm A + \diag(v_j)^{-1}\big)^{-1}$ and
  $\bm\mu_{s_j} = \sigma^{-2}\bm\Sigma_{s_j} \bm A^\top \bm x_j$,
  \[
    \bm s_j \mid \bm A,\sigma,\bm v,\bm\alpha,\bm\lambda
    \sim \NormRV\big(\bm s_j \mid \bm\mu_{s_j},\ \bm\Sigma_{s_j}\big),\quad j=1,\dots,N.
  \]
  \item Mixing matrix update (with Gaussian prior on $\bm A$):
  letting $\bm\Sigma_a = \sigma^2 \big(\sum_{j=1}^{N} \bm s_j \bm s_j^\top\big)^{-1}$ and
  $\bm\mu_{a} = \sigma^{-2}\bm\Sigma_a \sum_j \bm s_j \bm x_j^\top$,
  \[
    \bm A \mid \bm S,\sigma,\bm v,\bm\alpha,\bm\lambda
    \sim \NormRV\big(\bm A \mid \bm\mu_a,\ \bm\Sigma_a\big).
  \]
  \item Noise variance update:
  with $\gamma_{\sigma} = dN/2$ and $\beta_{\sigma} = 2 / \sum_j \|\bm x_j - \bm A \bm s_j\|_2^2$,
  \[
    \sigma \sim \sqrt{\InvGammaRV(\gamma_{\sigma}, \beta_{\sigma})}.
  \]
  \item Local scale update:
  let $\gamma_{v_i} = (\alpha_i + 1)/2$ and $\beta_{v_i,j} = 2/(s_{ij}^2 + \alpha_i \lambda_i^2)$, then
  \[
    v_{ij} \mid \bm S,\bm\alpha,\bm\lambda
    \sim \InvGammaRV\big(v_{ij} \mid \gamma_{v_i}, \beta_{v_i,j}\big).
  \]
  \item Hyperparameter updates for $\lambda_i$ (and possibly $\alpha_i$) from suitable gamma or other priors; see \citet{fevotte2004bayesian} for details.
\end{itemize}

\paragraph{Other source distributions.}
There have been a number of other source distributions analyzed in the literature. For example, finite mixtures of the form
\[
  p(\bm s) = \prod_{i=1}^S \sum_{\ell=1}^{M_i} p\big(s_i^{(\ell)}\big)\, p(\lambda_i=\ell),
\]
with $p(\lambda_i) \sim \text{Gamma}(r,c)$, appear in constrained EM algorithms for ICA \citep{hinton2001new}. All these examples fit into the same overarching Gaussian scale mixture picture. Our focus in this paper is on the \PG{}-augmented hyperbolic secantor horseshoe-type prior, for which the Gibbs sampler above yields a particularly clean and scalable Bayesian ICA algorithm.
\begin{algorithm}[t]
\caption{\PG-augmented Gibbs sampler for Bayesian ICA (\texttt{Gibbs-ICE})}
\label{alg:gibbs-ice}
\DontPrintSemicolon
\KwIn{Data matrix $\bm X \in \mathbb{R}^{N\times d}$, iterations $M$, noise variance $\sigma^2$, prior variance $\sigma_2^2$}
\KwOut{Posterior samples $\{(\bm S^{(m)}, \bm A^{(m)}, \bm T^{(m)})\}_{m=1}^M$}
\BlankLine

\textbf{Model:}
\begin{itemize}
  \item $\bm x_i \mid \bm s_i,\bm A \sim \NormRV_d(\bm A \bm s_i,\ \sigma^2 \bm I_d)$, $i=1,\dots,N$.
  \item $s_{ij} \mid \tau_{ij} \sim \NormRV\!\left(0,\frac{1}{4\tau_{ij}}\right)$, 
        $\tau_{ij} \sim PG(1,0)$, $i=1,\dots,N$, $j=1,\dots,d$.
  \item $\bm a_k \sim \NormRV_d(\bm 0,\ \sigma_2^2 \bm I_d)$, $k=1,\dots,d$.
\end{itemize}

\BlankLine
Initialize $\bm S^{(0)}, \bm A^{(0)}, \bm T^{(0)}$.\;
\For{$m = 1,\dots,M$}{
  \tcp{(1) Update sources $\bm S$}
  \For{$i = 1,\dots,N$}{
    Set $\bm D_i^{(m-1)} = \mathrm{diag}\big(4\tau_{i1}^{(m-1)},\dots,4\tau_{id}^{(m-1)}\big)$\;
    Compute
    \[
      \bm\Sigma_{s,i}^{(m)} = \Big(\sigma^{-2} (\bm A^{(m-1)})^\top \bm A^{(m-1)} + \bm D_i^{(m-1)}\Big)^{-1},
      \quad
      \bm\mu_{s,i}^{(m)} = \sigma^{-2} \bm\Sigma_{s,i}^{(m)} (\bm A^{(m-1)})^\top \bm x_i.
    \]
    Draw $\bm s_i^{(m)} \sim \NormRV_d\big(\bm\mu_{s,i}^{(m)}, \bm\Sigma_{s,i}^{(m)}\big)$.\;
  }
  Collect rows into $\bm S^{(m)} = (\bm s_1^{(m)},\dots,\bm s_N^{(m)})^\top$.\;

  \BlankLine
  \tcp{(2) Update mixing matrix $\bm A$}
  Compute 
  \[
    \bm\Sigma_a^{(m)} 
      = \Big(\sigma^{-2} (\bm S^{(m)})^\top \bm S^{(m)} + \sigma_2^{-2}\bm I_d\Big)^{-1}.
  \]
  \For{$k = 1,\dots,d$}{
    Let $\bm x_{\cdot k}$ be the $k$-th column of $\bm X$.\;
    Set $\bm\mu_{a,k}^{(m)} = \sigma^{-2}\bm\Sigma_a^{(m)} (\bm S^{(m)})^\top \bm x_{\cdot k}$.\;
    Draw $\bm a_k^{(m)} \sim \NormRV_d\big(\bm\mu_{a,k}^{(m)}, \bm\Sigma_a^{(m)}\big)$.\;
  }
  Collect columns into $\bm A^{(m)} = (\bm a_1^{(m)},\dots,\bm a_d^{(m)})$.\;

  \BlankLine
  \tcp{(3) Update \PG{} scales $\bm T$}
  \For{$i = 1,\dots,N$}{
    \For{$j = 1,\dots,d$}{
      Draw $\tau_{ij}^{(m)} \sim PG\big(1,\ \abs{2 s_{ij}^{(m)}}\big)$.\;
    }
  }
}
\end{algorithm}

\section{Theoretical results}\label{sec:theory}
Early Bayesian treatments of ICA used approximate inference owing to intractable posterior computation. For instance, \citet{attias1999independent}'s Independent Factor Analysis (IFA) presents a fully probabilistic ICA model with a flexible mixture of Gaussians prior on the sources and derives a variational EM algorithm for posterior inference. Several others adopted variational Bayes for ICA (e.g. \citet{Lappalainen1999,Choudrey2003}), attesting to the popularity of the approach. Though variational and mean-field methods develop efficient approximate inference schemes, they provide no guarantees on posterior concentration or the frequentist optimality of the Bayes estimator. Similarly, \citet{eloyan2013}, \citet{hyvarinen2002sparse} and related work use Bayesian formulations to encode sparsity or sign constraints on the mixing matrix, motivated by EEG and neural data, but focus mostly on computation and empirical performance, not asymptotics. 

\subsection*{Posterior Contraction for the Unmixing Matrix with Known Source Densities} 

In this work, we focus on the theoretical posterior contraction behavior for the ICA unmixing matrix under the simplifying assumption that the source distributions are known. Firstly, the known–source case serves as a simple benchmark, as it eliminates the complexity of simultaneously learning the source density shapes and allows us to isolate how well the Bayesian posterior learns the unmixing matrix $\bm{W}$ itself. If the posterior for the unmixing matrix can be shown to concentrate around the truth in this ideal scenario, it provides a baseline assurance that the Bayesian approach works under favorable conditions. Secondly, analyzing this case lays important groundwork for theoretical development in the realm of Bayesian ICA. 

The only general posterior contraction result we are aware of is due to
\citet{ShenEJS2016}, who study a nonparametric Bayesian block–ICA model.
They place priors on (i) the unknown block partition of the sources, (ii) the
mixing matrix, and (iii) the blockwise source densities using either random
series or Dirichlet mixture priors. Under suitable identifiability and
regularity conditions, they establish that the posterior for the joint
density $p$ of the latent sources contracts around the truth $p_0$ (up to
scale and permutation) at a minimax–optimal rate in Hellinger distance.
For classical ICA without block structure, their rate reduces to
$N^{-\alpha^\ast/(2\alpha^\ast+1)}$ (up to logarithmic factors), where
$\alpha^\ast$ is the worst marginal smoothness, matching the 1D nonparametric density rate. Their procedure is also adaptive to the unknown
smoothness and, in the block–ICA case, to the unknown block structure.
However, their analysis is purely on the density level and it does not provide a
$\sqrt{N}$ asymptotic normality result for the unmixing matrix itself. 

Our theoretical contribution is complementary to the rate–adaptive density
results of \citet{ShenEJS2016}. Whereas \citet{ShenEJS2016} treat both the
mixing matrix $\bm{A}=\bm{W}^{-1}$ and the blockwise source densities as unknown and
derive minimax–optimal contraction rates for the joint density of the latent
sources, we assume the marginal source densities $p_{0,k}$ are known and focus on the \emph{parametric} component of the ICA
model, namely the unmixing matrix $\bm{W}$. This allows us to work in a local
parametric neighborhood of $\bm{W}_0$, construct a uniform LAN expansion similar to \citet[Theorem~7.2]{vdv2000}, and
prove that the posterior for $\theta=\mathrm{vec}(\bm{W})$ contracts at the
parametric rate $N^{-1/2}$ and satisfies a Bernstein--von Mises theorem.
In particular, we obtain asymptotic normality of the posterior for $\bm{W}$
up to signed permutations, something that is not covered by the analysis of \citet{ShenEJS2016}. Our identifiability and loss function are
also different: we work with the signed–permutation distance $d_\pm$ on
$\mathrm{GL}(d)$ and show that $d_\pm$ is locally equivalent to the Frobenius
norm in a neighborhood of the canonical representative $\bm{W}_0$ (see Lemma~\ref{lem:local-equivalence}), so that the
posterior $\sqrt{N}$ concentration in $\theta$ translates directly to $\sqrt{N}$ posterior concentration in $d_\pm$.

\paragraph{Model and prior assumptions.}
Let $\bm{x}^{(1)},\dots,\bm{x}^{(N)}\in\mathbb{R}^d$ be i.i.d.\ observations from the Independent Component Analysis (ICA) model with density
\[
p_{\bm{W}}(x) = |\text{det}(\bm{W})|\prod_{k=1}^d p_{0,k}\!\big(\bm{w}_k^\top x\big),\quad \bm{W}\in\mathcal{W},
\]
where $\bm{w}_k^\top$ denotes the $k$th row of the unmixing matrix $\bm{W}$, the $p_{0,k}$ are the true source densities, and the parameter space $\mathcal{W}$ is an open subset of the general linear group $\mathrm{GL}(d)$.
We endow $\bm{W}$ with a prior $\Pi$ that has a continuous and strictly positive density with respect to the Lebesgue measure on $\mathrm{GL}(d)$ in a neighborhood of $\bm{W}_0$.
To measure the distance between an estimate $\bm{W}$ and the true class around $\bm{W}_0$, we define
\[
d_\pm(\bm{W},\bm{W}_0)=\inf_{D,P}\;\|\bm{W}-DP\bm{W}_0\|_F,
\]
where $\|\cdot\|_F$ is the Frobenius norm and the infimum is taken over all signed permutation matrices $DP$ (with $D$ diagonal with $\pm 1$ entries and $P$ a permutation matrix). This distance is natural because $\bm{W}_0$ is only identifiable up to signed permutations; distances between the raw matrices can be misleading, as two equally good solutions may differ by row permutations and sign flips.

We impose the following regularity and identifiability assumptions on the source densities $p_{0,k}$.

\begin{enumerate}
    \item[(A1)] \textbf{Smoothness and tails:} $p_{0,k}$ is strictly positive and three times continuously differentiable on $\mathbb{R}$. The score function $\psi_k(s)=\frac{d}{ds}\log p_{0,k}(s)$ satisfies $\lim_{|s|\to\infty} s\,p_{0,k}(s)=0$.
    \item[(A2)] \textbf{Moment conditions:} The source random variable $S_k \sim p_{0,k}$ satisfies $\E(S_k)=0$ and $\text{Var}(S_k)=1$. The score and its derivatives satisfy $\E\{\psi_k(S_k)^2\}<\infty$ and $\E\{|\psi'_k(S_k)|\}<\infty$.
    \item[(A3)] \textbf{Identifiability:} At most one source $p_{0,k}$ is Gaussian. For every non-Gaussian source, the Fisher information for location $J_k = \E\{\psi_k(S_k)^2\}$ is finite and positive. We also require non-singular Fisher information for the unmixing matrix, $I(\bm{W}_0) \succ 0$.
    \item[(A4)] \textbf{LAN regularity:} For the Local Asymptotic Normality (LAN) expansion to hold, we require, for some $\delta>0$, that the third derivative of the log-density is bounded in expectation:
    \[
      \E\Big[\sup_{|u|\le\delta} |\psi''_k(S_k+u)|\Big] < \infty 
      \quad \text{and} \quad 
      \E[|S_k|^3] < \infty.
    \]
    Additionally, we need a \emph{uniform domination} condition: there exists $\delta > 0$ and integrable envelopes $G_{1k},G_{2k}$ with $\E\{G_{1k}(S_k)\}< \infty$ and $\E\{G_{2k}(S_k)\|\bm S\|^3\}< \infty$ (where $\bm S=(S_1,\dots,S_d)$) such that 
    \[
      \sup_{|u| \le \delta} | \psi_{k}'(S_k+u)| \le G_{1k}(S_k), 
      \qquad 
      \sup_{|u| \le \delta} | \psi_k''(S_k + u)| \le G_{2k}(S_k).
    \]
\end{enumerate}
The $C^3$ regularity of $\log p_{0,k}$ and the integrability of the score and its derivatives in (A1)–(A2) ensure (i) differentiability under the integral sign, (ii) finite Fisher information $J_k=\E\{\psi_k(S_k)^2\}$ for the sources, and (iii) the uniform third-order bounds used in Lemma~\ref{lem:third-order}. Light moment conditions like $\E|S_k|^3<\infty$ are standard in parametric LAN analyses; see, e.g., \citet[Ch.~7]{vdv2000} and \citet{miller2021JMLR}. The ``at most one Gaussian'' condition in (A3) is the classical ICA identifiability guarantee \citep{comon1994independent,hyvarinen2001independent}. The non-singularity of the Fisher information for $\bm{W}$ at $\bm{W}_0$ is the standard curvature condition for BvM.  The envelopes $G_{1k},G_{2k}$ in (A4) provide a dominated convergence handle for the second and third derivatives of $\log p_{0,k}$ along the local paths $\bm{W}_0+t\Delta$. This is a routine strengthening needed to obtain \emph{uniform} LAN on growing neighborhoods; similar domination assumptions appear in parametric BvM proofs with matrix parameters, cf. \citet[Sec.~2.3]{vdv2000} and \citet{kleijnvdv2012}.

Next, we impose the following prior thickness and regularity condition near $\bm{W}_0$.
\begin{enumerate}
\item[(P1)] \textbf{Prior thickness:} There exist $\varepsilon_0>0$ and constants $0<c_1\le c_2<\infty$ such that the prior $\Pi$ admits a density
$\pi$ with respect to the Lebesgue measure on $\mathcal{W}$ and
\[
c_1 \le \pi(\bm{W}) \le c_2 \quad \text{for all } \bm{W} \text{ with } d_\pm(\bm{W},\bm{W}_0)<\varepsilon_0,
\]
and $\pi$ is continuous at $\bm{W}_0$.
\end{enumerate}
In the $\mathrm{GL}(d)$ case, $\Pi$ assigns no mass to singular matrices and has locally integrable tails. A continuous, strictly positive prior density on a neighborhood of $\bm{W}_0$ is the minimal thickness assumption in (P1) ensuring local flatness at $N^{-1/2}$ scales, which does not necessitate specific tails or global behavior in the parametric case \citep{castillonicklAOS2012}. Our choice of horseshoe-type priors slots in here, although their heavy-tailed nature is not needed specifically to prove contraction. 

\begin{theorem}[Posterior contraction and Bernstein-von Mises theorem for ICA] \label{thm:main}
Under the ICA model and assumptions \emph{(A1)--(A4)} and \emph{(P1)}, the posterior distribution for the unmixing matrix $\bm{W}$ concentrates at the parametric rate around the true signed-permutation class of $\bm{W}_0$. For any sequence $M_N\to\infty$,
\[
\Pi\!\left(\, d_\pm(\bm{W},\bm{W}_0) \;>\; \frac{M_N}{\sqrt N} \;\middle|\; \bm{x}^{(1)},\dots,\bm{x}^{(N)} \right)\;\overset{P_0}{\longrightarrow}\;0,
\]
where the convergence is in probability under the true data-generating process $P_0$.

Moreover, the posterior distribution is asymptotically Gaussian in the following sense:
let $I(\bm{W}_0)$ be the Fisher information matrix of the model at $\bm{W}_0$. The posterior for the vectorized matrix $\mathrm{vec}(\bm{W})$ satisfies a Bernstein--von Mises limit:
\[
\left\|\,\Pi\left(\,\sqrt N\{\mathrm{vec}(\bm{W})-\mathrm{vec}(\bm{W}_0)\}\in\cdot\ \middle|\ \bm{x}^{(1:N)}\right)
    \;-\; \mathcal{N}\left(\,\Delta_N,\ I(\bm{W}_0)^{-1}\right)\,\right\|_{\mathrm{TV}}
    \;\overset{P_0}{\longrightarrow}\;0,
\]
where $\Delta_N$ is an efficient estimator, such as the MLE, centered at the true value.
\end{theorem}

A sketch of the proof of Theorem~\ref{thm:main} follows the standard LAN-based route for finite-dimensional parametric Bayes procedures in \citet{miller2021JMLR,Bickel_2012}. The key ingredient in proving Theorem~\ref{thm:main} is the following Uniform Local Asymptotic Normality (ULAN) lemma, which gives a quadratic expansion of the log-likelihood around the true parameter $\theta_0$ for local perturbations $h$ in a fixed ball. 

\begin{lemma}[Uniform Local Asymptotic Normality]\label{lem:LAN}
Let $\theta=\mathrm{vec}(\bm{W})\in\mathbb{R}^{d^2}$, $\theta_0=\mathrm{vec}(\bm{W}_0)$, and $h=\sqrt{N}\,(\theta-\theta_0)$. Define the per-sample log-likelihood $\ell(\bm{W};\bm{x})=\log p_{\bm{W}}(\bm{x})$ and the total log-likelihood
\[
L_N(\theta)=\sum_{n=1}^N \ell\big(\bm{W}(\theta);\bm{x}^{(n)}\big),
\]
where $\bm{W}(\theta)$ denotes the $d\times d$ matrix obtained by reshaping $\theta$.
Under \emph{(A1)--(A4)}, for any fixed $R>0$,
\[
\sup_{\|h\|\le R}\ \Big|\,
L_N(\theta_0+h/\sqrt N)-L_N(\theta_0)-h^\top S_N+\tfrac{1}{2}h^\top \mathcal I h
\,\Big|\ \xrightarrow{P_0}\ 0,
\]
where
\[
S_N \;:=\; \frac{1}{\sqrt N}\sum_{n=1}^N \nabla_\theta \ell\big(\bm{W}_0;\bm{x}^{(n)}\big) 
\ \Rightarrow\ \mathcal N(0,\mathcal I),
\]
and $\mathcal I=-\E\big[\nabla_\theta^2 \ell(\bm{W}_0;X)\big]=I(\bm{W}_0)$ is the per-observation Fisher information.
\end{lemma}

The proof of Lemma~\ref{lem:LAN} itself is built from three pieces:
(i) the integration-by-parts identities in Lemma~\ref{lem:IBP-identities}, which show that the score has mean zero and identify the information matrix; (ii) the third-order Taylor remainder bound in Lemma~\ref{lem:third-order}, which uses the envelope conditions in (A4) to control $D^3 \ell(\bm{W};X)$ uniformly along local paths $\bm{W}_0 + t \Delta$; and (iii) LLNs for the Hessian to ensure $-N^{-1} \nabla_{\theta}^2 L_N(\theta_0) \rightarrow \mathcal{I}.$ Assumption (P1) is then used in Lemma~\ref{lem:prior} to show the prior is locally flat on the $N^{-1/2}$ scale, so the posterior kernel in the $h$-parametrization is essentially
\[
q_{N}(h) \;\propto\; \exp\!\Big(h^\top S_N - \tfrac{1}{2}h^\top \mathcal{I} h + r_N (h)\Big),
\]
up to a constant factor $\pi(\bm{W}_0)$. Lemma~\ref{lem:normalizers} compares the normalizing constant $Z_N=\int q_N(h)\,dh$ to the Gaussian normalizer $Z_N^0$, showing $Z_N / (\pi(\bm{W}_0)Z_N^0)\to 1$ in $P_0$–probability. This yields that the posterior for $h$ is asymptotically close in total variation to $\mathcal N(\mathcal I^{-1}S_N,\mathcal I^{-1})$; the BvM part of Theorem~\ref{thm:main} is just this statement rewritten in terms of $\theta=\mathrm{vec}(\bm{W})$. Finally, for the contraction part, one uses the same quadratic bound on the exponent to show that the posterior mass outside balls $\{ \|h\| \le M_N\}$ with $M_N\to\infty$ vanishes, and then invokes the local equivalence between $d_\pm(\bm{W},\bm{W}_0)$ and the Euclidean norm $\|\theta-\theta_0\|$ from Lemma~\ref{lem:local-equivalence}. Together, these steps establish both $\sqrt{N}$ contraction in the identified metric $d_\pm$ and the Bernstein--von Mises limit.

\paragraph{On the choice and growth of $M_N$.}
Theorem~\ref{thm:main} is stated with an arbitrary divergent sequence $M_N\to\infty$.
In practice one may take $M_N=\log\log N$, or $M_N=(\log N)^\alpha$ with $\alpha>0$. The proof only requires that the LAN remainder and uniform Hessian LLN hold on balls of radius $R_N$ with $M_N=o(R_N)$ (we use $R_N=N^{1/6-\eta}$). The rate $1/\sqrt N$ is sharp (parametric), and $M_N$ simply modulates how far into the tails we ask the posterior mass to vanish.

\section{Applications}\label{sec:app}
\subsection{Numerical Experiments}\label{sec:toyexample}

We begin with a small, controlled experiment to verify that our Gibbs sampler
recovers sources and loadings under a generative model consistent with our
augmentation. Throughout this subsection we fix \(n=500\) observations and
\(d=4\) components, and generate data according to
\begin{align}
\bm{V} &\sim \mathcal{N}_{d\times d}(\bm{0},\sigma_2^2 I_d), \nonumber\\
\tau_{ij} &\stackrel{\text{iid}}{\sim} \operatorname{PG}(1,0),
\quad i=1,\dots,n,\; j=1,\dots,d, \nonumber\\
s_{ij}\mid \tau_{ij} &\sim \mathcal{N}\bigl(0,(4\tau_{ij})^{-1}\bigr), \nonumber\\
\bm{x}_i \mid (\bm{s}_i, \bm{V}) &\sim \mathcal{N}_d\bigl(\bm{V}^\top \bm{s}_i,\ \sigma^2 I_d\bigr),
\quad i=1,\dots,n, \label{eq:dgp}
\end{align}
so that \(\bm{X} = \bm{S}\bm{V}^\top + \bm{\varepsilon}\) with
\(\bm{\varepsilon} \sim \mathcal{N}_{n\times d}(\bm{0},\sigma^2 I)\).
We run our \texttt{Gibbs-ICE} sampler on \(\bm{X}\) and compare posterior summaries to the
ground truth \(\bm{S}\). For visualization, we apply a post-processing alignment
that fixes permutation and sign.

\begin{figure}[h]
  \centering
  \includegraphics[width=0.6\linewidth]{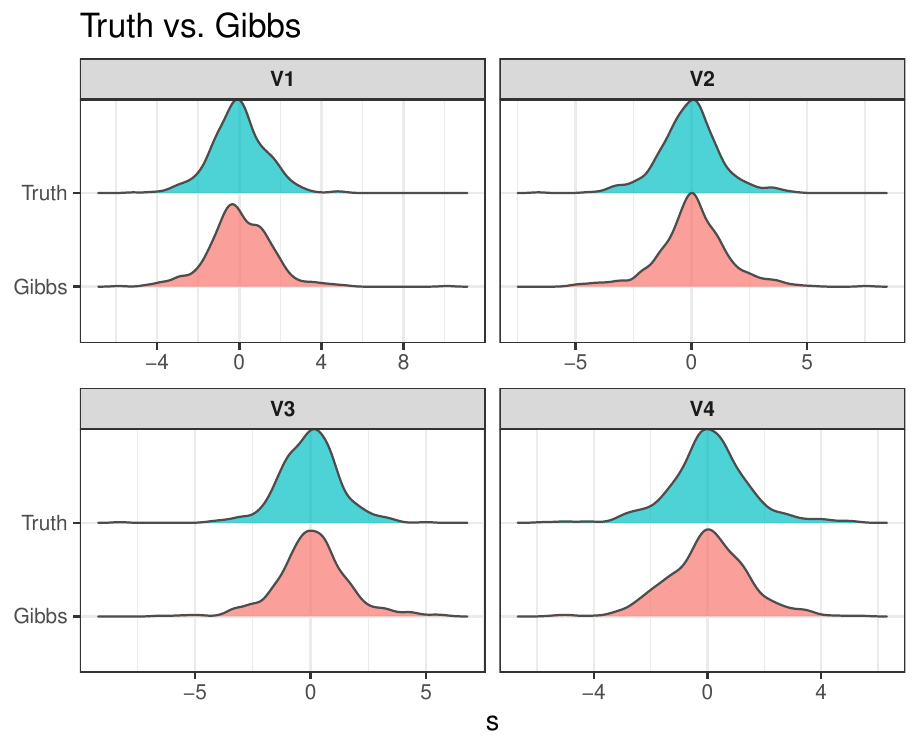}
  \caption{Posterior vs. true source densities (low noise).}
  \label{fig:ridge-gibbs-case1}
\end{figure}

\paragraph{Case 1: Well-separated signals, low noise.}
We set \(\sigma=0.01\) and \(\sigma_2=1\).
Figure~\ref{fig:ridge-gibbs-case1} overlays the marginal densities of each
recovered source \(\hat{s}_{\cdot j}\) (posterior mean) against the
corresponding true \(s_{\cdot j}\).
The Gibbs sampler closely tracks all four components: the peaks and tail
thickness align well, and there is no visible mode splitting or spurious
heavy-tailing. This indicates that (i) the conditional Gaussian updates for
\(\bm{S}\) given \(\bm{\tau}\) and \(\bm{V}\) are numerically stable, and
(ii) the overall mixing across sign and scale ambiguities has been handled
adequately via alignment.

\begin{figure}[h]
  \centering
  \includegraphics[width=0.6\linewidth]{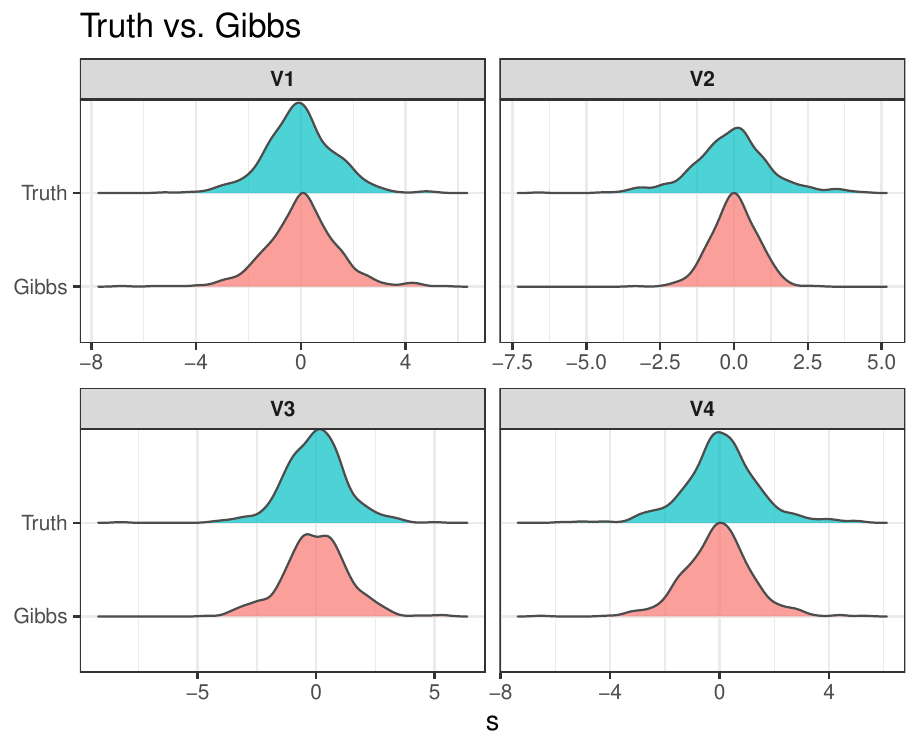}
  \caption{Posterior vs. true source densities, Case 2 (one spiky component and higher noise).}
  \label{fig:ridge-gibbs-case2}
\end{figure}

\paragraph{Case 2: One ``hard'' component and higher noise.}
We increase difficulty by (a) scaling the first \(\tau_{\cdot 1}\) by \(100\) times, making the first source substantially more peaked near \(0\), and
(b) raising the observation noise to \(\sigma=0.1\) in the data-generating
process~\eqref{eq:dgp}.
Figure~\ref{fig:ridge-gibbs-case2} shows the same density overlays.
As expected, the first component becomes challenging: \texttt{Gibbs-ICE} still centers
the mass correctly but shows attenuation near the spike and slightly broader
shoulders, reflecting genuine non-identifiability introduced by a
near-degenerate source convolved with higher noise.
The remaining three components remain well recovered, with shapes and tail
behavior close to truth.

When the latent scales \(\tau\) and the noise level are benign (Case 1), \texttt{Gibbs-ICE} recovers the univariate margins of each source very closely, indicating that the conditional P\'olya–Gamma blocks behave as intended. With a deliberately low variance component and higher observation noise (Case 2), the first source becomes harder, while the other components remain stable: matching
intuition about ICA identifiability under super-Gaussian priors.

\subsection{Extensive simulations against contemporary ICA methods}\label{sec:contemporarymethods}

We benchmark \texttt{Gibbs-ICE} against widely used ICA estimators:
\texttt{FastICA} (fixed-point negentropy maximization; \citealp{Hyvarinen2000}),
\texttt{ProDenICA} (projection pursuit with nonparametric density estimation; \citealp{HastieNIPS}),
\texttt{JADE} (joint approximate diagonalization of fourth-order cumulants; \citealp{NordhausenGazen2022}),
\texttt{SOBI} (second-order blind identification via joint diagonalization of time-lagged covariances; \citealp{Belouchrani1998BlindSS}),
\texttt{FOBI} (fourth-order blind identification using kurtosis; \citealp{Miettinen2015}),
\texttt{PearsonICA} (contrast based on Pearson divergences; \citealp{Karvanen2002}),
and \texttt{steadyICA} (stable $M$-estimators with robust scatter; \citealp{matteson2013ICA}).
Unless stated otherwise, competing methods receive the same centered or whitened inputs and use their default contrast functions.

\paragraph{Evaluation metrics.}
Because ICA is identifiable only up to permutation and sign, we first align estimated sources \(\hat{\bm{S}}\) to the truth \(\bm{S}\) by solving a linear assignment on absolute correlations (Hungarian algorithm; \citealp{kuhn1955}).
We then report:
(i) the \emph{Amari distance} between the estimated unmixing \(\hat{\bm{W}}\) and the true \(\bm{W}\),
\[
d_{\text{Amari}}(\hat{\bm{W}},\bm{W})
=\frac{1}{2(d-1)}
\Bigg[
\sum_{i=1}^d \Big(\frac{\sum_j |A_{ij}|}{\max_j |A_{ij}|}-1\Big)
+
\sum_{j=1}^d \Big(\frac{\sum_i |A_{ij}|}{\max_i |A_{ij}|}-1\Big)
\Bigg],\quad \bm{A}=\bm{W}^{-1}\hat{\bm{W}},
\]
a sign- and permutation–invariant matrix error widely used in ICA
(\citealp{Amari1963}); 
(ii) the \emph{source recovery correlation} (SRC), defined as the mean diagonal of \(\big|\text{corr}(\hat{\bm{S}}_{\text{aligned}},\bm{S})\big|\) after alignment; and 
(iii) the \emph{reconstruction RMSE},
\[
\text{RMSE} \;=\; \frac{\big\|\bm{X}-\hat{\bm{S}}_{\text{aligned}}\hat{\bm{A}}^\top\big\|_F}{\sqrt{nd}},
\quad \hat{\bm{A}}=\hat{\bm{W}}^{-1}.
\]
Lower Amari and RMSE and higher SRC indicate better performance.

We consider four super-Gaussian source families commonly used in ICA stress tests:
(i) \textbf{sech}, (ii) \textbf{Student-$t_3$},
(iii) \textbf{Laplace}, and
(iv) \textbf{mixed} (half $t_3$, half Laplace). For each scenario we set \((n,d)\in\{(500,4),(2000,8)\}\) and noise \(\sigma\in\{0.01,0.05\}\), draw \(\bm{S}\) i.i.d.\ from the chosen family and \(\bm{X} = \bm{S}\bm{A}^\top + \bm{\varepsilon}\) with a well-conditioned random \(\bm{A}\) and \(\bm{\varepsilon}\sim\mathcal N(\bm{0},\sigma^2 I)\).
All methods are run with their recommended defaults; \texttt{Gibbs-ICE} uses \(4{,}000\) iterations (first \(2{,}000\) burn-in), thinning by \(5\), and posterior means for summaries.

\begin{table}[h]
\centering
\caption{Aggregate ICA performance by source family. Lower Amari/RMSE and higher SRC are better. Best in each column within a source family is bolded.}
\label{tab:ica-agg}
\setlength{\tabcolsep}{6pt}
\begin{tabular}{l l c c c}
\toprule
\multirow{2}{*}{\textbf{Method}} & 
\multirow{2}{*}{\textbf{Source family}} &
\multicolumn{3}{c}{\textbf{Performance}} \\
\cmidrule(lr){3-5}
 & & \textbf{Amari} $\downarrow$ & \textbf{SRC} $\uparrow$ & \textbf{RMSE} $\downarrow$ \\
\midrule
\texttt{FOBI}              & \multirow{8}{*}{\hspace{2em} \textbf{sech}}   & 2.021 & 0.847 & 3.024 \\
\texttt{JADE}              &                                   & 1.691 & \textbf{0.986} & \textbf{2.497} \\
\texttt{SOBI}              &                                   & 1.810 & 0.806 & 3.471 \\
\texttt{FastICA}           &                                   & 2.005 & 0.960 & 2.777 \\
\texttt{PearsonICA}        &                                   & 1.975 & 0.725 & 2.882 \\
\texttt{steadyICA}         &                                   & 1.477 & 0.863 & 3.383 \\
\texttt{ProDenICA}         &                                   & \textbf{1.423} & 0.851 & 2.589 \\
\texttt{Gibbs-ICE}         &                                   & 1.787 & 0.858 & 2.577 \\
\midrule
\texttt{FOBI}              & \multirow{8}{*}{\hspace{2em} $\mathbf{t}_3$}  & 1.530 & 0.872 & 2.251 \\
\texttt{JADE}              &                                   & 1.397 & 0.988 & 2.396 \\
\texttt{SOBI}              &                                   & 1.510 & 0.739 & 3.458 \\
\texttt{FastICA}           &                                   & \textbf{1.203} & \textbf{0.994} & \textbf{2.096} \\
\texttt{PearsonICA}        &                                   & 1.577 & 0.772 & 3.471 \\
\texttt{steadyICA}         &                                   & 1.717 & 0.991 & 2.561 \\
\texttt{ProDenICA}         &                                   & 1.782 & 0.798 & 3.396 \\
\texttt{Gibbs-ICE}         &                                   & 1.480 & 0.725 & 2.546 \\
\midrule
\texttt{FOBI}              & \multirow{8}{*}{\hspace{2em} \textbf{Laplace}}& 1.549 & 0.895 & 3.145 \\
\texttt{JADE}              &                                   & \textbf{1.263} & 0.992 & 3.151 \\
\texttt{SOBI}              &                                   & 1.449 & 0.748 & 3.027 \\
\texttt{FastICA}           &                                   & 1.693 & \textbf{0.993} & 3.432 \\
\texttt{PearsonICA}        &                                   & 1.610 & 0.830 & 3.602 \\
\texttt{steadyICA}         &                                   & 1.595 & 0.864 & 3.520 \\
\texttt{ProDenICA}         &                                   & 1.521 & 0.719 & 3.784 \\
\texttt{Gibbs-ICE}         &                                   & 1.543 & 0.983 & \textbf{2.444} \\
\midrule
\texttt{FOBI}              & \multirow{8}{*}{\hspace{2em} \textbf{Mixed}}  & 1.536 & 0.882 & 2.420 \\
\texttt{JADE}              &                                   & \textbf{1.421} & 0.981 & 2.414 \\
\texttt{SOBI}              &                                   & 1.844 & 0.793 & 2.854 \\
\texttt{FastICA}           &                                   & 1.854 & \textbf{0.9973} & 3.079 \\
\texttt{PearsonICA}        &                                   & 2.148 & 0.685 & \textbf{1.354} \\
\texttt{steadyICA}         &                                   & 1.695 & 0.992 & 3.168 \\
\texttt{ProDenICA}         &                                   & 1.833 & 0.883 & 2.759 \\
\texttt{Gibbs-ICE}         &                                   & 2.129 & 0.740 & 2.217 \\
\bottomrule
\end{tabular}
\end{table}

Table~\ref{tab:ica-agg} shows that across super-Gaussian \textbf{sech} sources, \texttt{JADE} achieves the top SRC and the lowest RMSE, while \texttt{ProDenICA} attains the best (lowest) Amari distance. For heavy-tailed $t_3$ sources, \texttt{FastICA} dominates all three metrics. With Laplace sources, \texttt{JADE} gives the best Amari, \texttt{FastICA} yields the highest SRC, and \texttt{Gibbs-ICE} delivers the best reconstruction error (lowest RMSE), indicating strong mixing-matrix recovery up to permutation. In the mixed family, \texttt{JADE} minimizes Amari, \texttt{FastICA} maximizes SRC, and \texttt{PearsonICA} attains the lowest RMSE, suggesting that different methods excel under different criteria and distributional regimes.

Thus, if the downstream goal emphasizes accurate reconstruction or denoising, \texttt{Gibbs-ICE} is particularly attractive, especially with super-Gaussian but not ultra heavy tails. However, if the priority is pure separation under very heavy tails, contrast-based ICA methods like \texttt{FastICA} tend to win on Amari distance or SRC.

\section{Discussion}\label{sec:discuss}
High dimensional feature extraction has been a central tool in many modern applications.
Pattern–matching methods such as deep learning compose nonlinear maps to find features. However, nonlinear ICA is notoriously hard to learn, and recent work has combined flow–based models with ICA.
We develop a unified framework for optimization and posterior simulation using auxiliary variable methods. The key device is a Gaussian scale–mixture representation of the MacKay source prior
$1/\cosh(s)$ with a P\'olya–Gamma mixing density, which yields conditionally Gaussian updates for the
sources and loadings. This representation makes it possible to implement a fully Bayesian blockwise Gibbs sampler (see Algorithm \ref{alg:gibbs-ice}). The resulting view also connects classic contrast–based ICA to envelope
optimization outlined in Appendix \ref{sec:env}, and it links directly to the Bayesian ICA literature that
employs $t$, Laplace, and Jeffreys type source priors through a common scale–mixture lens
\citep{mackay1992bayesian, fevotte2004bayesian}. 
\paragraph{Toward semiparametric posterior contraction.}
Beyond our parametric theory for $W$ with known source densities, we are developing semiparametric
posterior contraction results in which the unmixing matrix $\bm{W}$ is finite-dimensional, but the
marginal source densities are \emph{unknown} and modeled nonparametrically. The goal is to
obtain joint rates in a metric that respects ICA identifiability (e.g., $d_{\pm}$ for $\bm{W}$ and
Hellinger for the product density of the latent sources), with
(i) a parametric $\sqrt{N}$ rate for $\bm{W}$ and
(ii) nonparametric rates $N^{-\alpha/(2\alpha+1)}$ (up to log factors) for the source densities, governed by
their marginal smoothness $\alpha$ as in density estimation. In this setting, horseshoe–like priors
on the sources are expected to play a more pronounced role: their global–local structure enforces strong shrinkage toward sparse or near–sparse coordinates while preserving \emph{ultra heavy tails},
which helps stabilize the estimation of $\bm{W}$ by adapting to unknown tail behavior and local regularity of the source densities without hand tuned
bandwidths. Technically, our analysis follows a semiparametric route: we construct score functions
orthogonal to the nuisance tangent space to obtain a uniform LAN expansion for $\bm{W}$, verify prior
thickness and small–ball probabilities for the source priors, and control entropy of suitable sieves
for the latent densities, in the spirit of adaptive nonparametric Bayes. This analysis complements the
density level contraction of \citet{ShenEJS2016}, who treat both mixing and densities as unknown, by
delivering parametric inference for $W$ and nonparametric learning of the sources within a single
Bayesian procedure driven by horseshoe–type shrinkage.

\bibliographystyle{plainnat}
\bibliography{ica-hs-review, glref, ref}
\newpage
\appendix

\section{Additional Theoretical Results}
\label{sec:additional_theory}
We need the following identities to prove our main theoretical results.
\begin{lemma}\label{lem:IBP-identities}
Under {\rm(A1)}--{\rm(A2)}, the following identities hold:
\begin{align}
\mathbb{E}\big[\psi_k(S_k)\,S_k\big] &= -1, \qquad k=1,\dots,d, \label{eq:psiSkSk}\\
\mathbb{E}\big[\psi(\bm{S})\,\bm{S}^\top\big] &= -I_d, \label{eq:psiSS}
\end{align}
and consequently, with $\bm{A}_0=\bm{W}_0^{-1}$ and $\bm{X}=\bm{A}_0\bm{S}$,
\begin{equation}\label{eq:EU0}
\mathbb{E}\,U(\bm{W}_0;\bm{X})
\;=\; (\bm{W}_0^{-1})^\top \;+\; \mathbb{E}\!\big[\psi(\bm{S})\,\bm{X}^\top\big]
\;=\; (\bm{W}_0^{-1})^\top \;-\; \bm{A}_0^\top \;=\; 0.
\end{equation}
\end{lemma}

\begin{proof}
Fix $k\in\{1,\dots,d\}$. By definition $\psi_k(s)=\frac{d}{ds}\log p_{0,k}(s)=\frac{p'_{0,k}(s)}{p_{0,k}(s)}$, so
\[
\mathbb{E}\big[\psi_k(S_k)\,S_k\big]
= \int_{\mathbb{R}} \psi_k(s)\,s\,p_{0,k}(s)\,ds
= \int_{\mathbb{R}} s\,p_{0,k}'(s)\,ds.
\]
Apply integration by parts with $u(s)=s$ and $dv=p'_{0,k}(s)\,ds$ (hence $du=ds$ and $v=p_{0,k}(s)$):
\begin{align*}
\int_{\mathbb{R}} s\,p_{0,k}'(s)\,ds
&= \big[s\,p_{0,k}(s)\big]_{-\infty}^{+\infty} \;-\; \int_{\mathbb{R}} p_{0,k}(s)\,ds.
\end{align*}
Assumption {\rm(A1)} gives $\lim_{|s|\to\infty} s\,p_{0,k}(s)=0$, so the boundary term vanishes, and since $p_{0,k}$ is a density,
$\int_{\mathbb{R}} p_{0,k}(s)\,ds=1$. Therefore \eqref{eq:psiSkSk} holds:
\[
\mathbb{E}\big[\psi_k(S_k)\,S_k\big] \;=\; 0 - 1 \;=\; -1.
\]

Next, write $\bm{S}=(S_1,\dots,S_d)^\top$ with independent coordinates and joint density $\prod_{m=1}^d p_{0,m}(s_m)$.
The $(k,j)$ entry of $\mathbb{E}[\psi(\bm{S})\bm{S}^\top]$ is
\begin{align*}
\mathbb{E}\big[\psi_k(S_k)\,S_j\big]
&= \int_{\mathbb{R}^d} \psi_k(s_k)\, s_j \,\prod_{m=1}^d p_{0,m}(s_m)\, d s_1\cdots d s_d.
\end{align*}
If $j=k$, Fubini reduces this to the one–dimensional identity \eqref{eq:psiSkSk}, giving
$\mathbb{E}[\psi_k(S_k)S_k]=-1$.
If $j\neq k$, independence factorizes the integral:
\begin{align*}
\mathbb{E}\big[\psi_k(S_k)\,S_j\big]
&= \Big(\int_{\mathbb{R}}\psi_k(s_k)\,p_{0,k}(s_k)\,ds_k\Big)\,
   \Big(\int_{\mathbb{R}} s_j\,p_{0,j}(s_j)\,ds_j\Big)
= \mathbb{E}\big[\psi_k(S_k)\big]\cdot \mathbb{E}[S_j].
\end{align*}
Now
\[
\mathbb{E}\big[\psi_k(S_k)\big]
= \int_{\mathbb{R}} \frac{p'_{0,k}(s)}{p_{0,k}(s)}\,p_{0,k}(s)\,ds
= \int_{\mathbb{R}} p'_{0,k}(s)\,ds
= \big[p_{0,k}(s)\big]_{-\infty}^{+\infty} \;=\; 0,
\]
since {\rm(A1)} implies $p_{0,k}(s)\to 0$ as $|s|\to\infty$ (indeed $s\,p_{0,k}(s)\to 0$ implies $p_{0,k}(s)=o(1/|s|)$).
Therefore $\mathbb{E}[\psi_k(S_k)S_j]=0$ for $j\neq k$. Combining the diagonal and off–diagonal cases yields \eqref{eq:psiSS}:
\[
\mathbb{E}\big[\psi(\bm{S})\,\bm{S}^\top\big] \;=\; -I_d .
\]

Finally, with $\bm{X}=\bm{A}_0\bm{S}$ and $\bm{A}_0=\bm{W}_0^{-1}$, compute the mean of the matrix score
\[
U(\bm{W}_0;\bm{X}) \;=\; (\bm{W}_0^{-1})^\top \;+\; \psi(\bm{S})\,\bm{X}^\top \;=\; (\bm{W}_0^{-1})^\top \;+\; \psi(\bm{S})\,\bm{S}^\top \bm{A}_0^\top,
\]
to obtain
\[
\mathbb{E}\,U(\bm{W}_0;\bm{X})
= (\bm{W}_0^{-1})^\top \;+\; \mathbb{E}[\psi(\bm{S})\bm{S}^\top]\,\bm{A}_0^\top
= (\bm{W}_0^{-1})^\top \;-\; \bm{A}_0^\top \;=\; 0,
\]
which is \eqref{eq:EU0}.
\end{proof}

\begin{lemma}[Third--order Taylor remainder]\label{lem:third-order}
Let $\theta=\mathrm{vec}(\bm{W})\in\mathbb{R}^{p}$, $\theta_0=\mathrm{vec}(\bm{W}_0)$, and $h=\sqrt N\,(\theta-\theta_0)$.
Write $\Delta:=\frac{1}{\sqrt N}\,\mathrm{mat}(h)\in\mathbb{R}^{d\times d}$ and $\bm{W}_t:=\bm{W}_0+t\,\Delta$ for $t\in[0,1]$.
For one observation $\bm{X}$, set
\[
\ell(\bm{W};\bm{X}) \;=\; \log|\det \bm{W}| \;+\; \sum_{k=1}^d \log p_{0,k}(\bm{w}_k^\top \bm{X}),
\]
where $\bm{w}_k^\top$ is the $k$th row of $\bm{W}$.
Assume {\rm(A1)--(A4)}. Then for any fixed $R>0$, there exists $N_0(R)$ such that for all $N\ge N_0(R)$,
the third--order Taylor remainder of the $N$-sample log-likelihood,
\[
R_N(h)\ :=\ \sum_{i=1}^N \int_0^1 \frac{(1-t)^2}{2}\; D^3 \ell(\bm{W}_t;\bm{X}^{(i)})\,[\Delta,\Delta,\Delta]\;dt,
\]
satisfies
\[
\sup_{\|h\|\le R}\ |R_N(h)| \ =\ O_P\!\Big(\frac{\|h\|^3}{\sqrt N}\Big).
\]
\end{lemma}

\begin{proof}
Throughout, $\|\cdot\|$ denotes the Euclidean norm for vectors and the Frobenius norm for matrices; $\|\cdot\|_{\mathrm{op}}$ is the operator norm. Since the map $\theta\mapsto\bm{W}$ is linear, there exists a constant $c_T>0$ such that
\begin{equation}\label{eq:Delta-norm}
\|\Delta\| \ \le\ \frac{c_T}{\sqrt N}\,\|h\|.
\end{equation}

\paragraph{Uniform invertibility along the path.}
Fix $R>0$. Because $\bm{W}_0$ is nonsingular and the inversion map $\bm{W}\mapsto \bm{W}^{-1}$ is continuous, there exists $\eta>0$ and a finite constant $C_\ast$ such that
\[
\|\bm{W}-\bm{W}_0\|_F<\eta\quad\Longrightarrow\quad \|\bm{W}^{-1}\|_{\mathrm{op}} \le C_\ast.
\]
Choose $N_0(R)$ so large that $\sup_{\|h\|\le R}\|\Delta\|\le \eta$ for all $N\ge N_0(R)$, cf.\ \eqref{eq:Delta-norm}. Then for all $t\in[0,1]$, $N\ge N_0(R)$ and $\|h\|\le R$,
\begin{equation}\label{eq:W-inv-bound}
\|\bm{W}_t^{-1}\|_{\mathrm{op}} \ \le\ C_\ast
\quad\text{and hence}\quad
\|\bm{W}_t^{-1}\|_{\mathrm{op}}^3 \ \le\ C_\ast^3 .
\end{equation}

\paragraph{Third order integral remainder and decomposition.}
We use the third--order integral remainder for each observation $\bm{X}$:
\[
R^{(1)}(h;\bm{X})\ :=\ \int_0^1 \frac{(1-t)^2}{2}\; D^3 \ell(\bm{W}_t;\bm{X})\,[\Delta,\Delta,\Delta]\;dt,
\qquad R_N(h) \;=\; \sum_{i=1}^N R^{(1)}(h;\bm{X}^{(i)}).
\]
We bound separately the contributions from $\log|\det \bm{W}|$ and from the source likelihood $\ell_{\textrm{src}}=\sum_{k=1}^d\log p_{0,k}(\bm{w}_k^\top \bm{X})$.

Matrix differential calculus yields
\[
D\,\log|\det \bm{W}|\,[H]=\mathrm{tr}(\bm{W}^{-1}H),\quad
D^2\,\log|\det \bm{W}|\,[H,K] = -\,\mathrm{tr}(\bm{W}^{-1}H\bm{W}^{-1}K),
\]
and
\[
D^3\,\log|\det \bm{W}|\,[H,H,H] \;=\; 2\,\mathrm{tr}\big(\bm{W}^{-1}H\bm{W}^{-1}H\bm{W}^{-1}H\big).
\]
Hence, with $H=\Delta$ and $\bm{W}=\bm{W}_t$,
\[
\big|D^3\,\log|\det \bm{W}_t|\,[\Delta,\Delta,\Delta]\big|
\;\le\; 2\,\|\bm{W}_t^{-1}\|_{\mathrm{op}}^3\,\|\Delta\|^3
\;\le\; 2\,C_\ast^3\,\|\Delta\|^3,
\]
by \eqref{eq:W-inv-bound}. Therefore, for each $\bm{X}$,
\[
\left|\int_0^1 \frac{(1-t)^2}{2}\,D^3 \log|\det \bm{W}_t|\,[\Delta,\Delta,\Delta]\,dt\right|
\ \le\ \frac{2\,C_\ast^3}{2}\,\|\Delta\|^3
\ \le\ \frac{C_1}{N^{3/2}}\,\|h\|^3
\]
with $C_1:=C_\ast^3 c_T^3$. Summing over $i=1,\dots,N$ gives
\begin{equation}\label{eq:det-bound}
\sum_{i=1}^N \left|\int_0^1 \frac{(1-t)^2}{2}\,D^3 \log|\det \bm{W}_t|\,[\Delta,\Delta,\Delta]\,dt\right|
\ \le\ N\,\frac{C_1}{N^{3/2}}\,\|h\|^3
\ =\ \frac{C_1}{\sqrt N}\,\|h\|^3 .
\end{equation}

Let $s_k(\bm{W}):=\bm{w}_k^\top \bm{X}$. Then $D s_k(\bm{W})[H]=H_k \bm{X}$ and $D^2 s_k\equiv 0$.
With $\psi_k=\partial_s\log p_{0,k}$ and $\psi_k',\psi_k''$ its derivatives,
\begin{align*}
D\,\Big(\sum_{k=1}^d \log p_{0,k}(s_k(\bm{W}))\Big)[H] & = \sum_{k=1}^d \psi_k(s_k(\bm{W}))\,(H_k \bm{X}),\\
D^2\,\Big(\sum_{k=1}^d \log p_{0,k}(s_k(\bm{W}))\Big)[H,H] & = \sum_{k=1}^d \psi_k'(s_k(\bm{W}))\,(H_k \bm{X})^2,\\
D^3\,\Big(\sum_{k=1}^d \log p_{0,k}(s_k(\bm{W}))\Big)[H,H,H] & = \sum_{k=1}^d \psi_k''(s_k(\bm{W}))\,(H_k \bm{X})^3.
\end{align*}
Therefore, at $\bm{W}_t$ and $H=\Delta$,
\begin{equation}\label{eq:src-D3-pointwise}
\big|D^3\,\ell_{\rm src}(\bm{W}_t;\bm{X})\,[\Delta,\Delta,\Delta]\big|
\ \le\ \sum_{k=1}^d \sup_{u\in \mathcal U_{k,t}(\bm{X})} |\psi_k''(u)|\,|\Delta_k \bm{X}|^3,
\end{equation}
where $s_k(\bm{W}_t)=\bm{w}_{0,k}^\top \bm{X} + t\,\Delta_k \bm{X} =: S_k + t\,\Delta_k \bm{X}$, hence
\[
\mathcal U_{k,t}(\bm{X})\ :=\ \{\,S_k + \tau\,\Delta_k \bm{X}:\ \tau\in[0,1]\,\}.
\]

Assumption (A4) implies in particular $\mathbb{E} G_{2k}(S_k)<\infty$ and $\mathbb{E}\|\bm{X}\|^3\sum_k G_{2k}(S_k)<\infty$ since $\bm{X}=\bm{A}_0\bm{S}$.

\medskip
\noindent Define the event
\[
\mathcal{M}(\bm{X})\ :=\ \Big\{\ \max_{1\le k\le d}\, |\Delta_k \bm{X}|\ \le\ \delta\ \Big\}.
\]
On $\mathcal{M}(\bm{X})$, (A4) and \eqref{eq:src-D3-pointwise} give
\[
\sup_{t\in[0,1]}\big|D^3\,\ell_{\rm src}(\bm{W}_t;\bm{X})\,[\Delta,\Delta,\Delta]\big|
\ \le\ \sum_{k=1}^d G_{2k}(S_k)\,|\Delta_k \bm{X}|^3
\ \le\ \|\Delta\|^3\,\|\bm{X}\|^3\,\sum_{k=1}^d G_{2k}(S_k),
\]
and hence
\begin{equation}\label{eq:src-oneobs}
\int_0^1 \frac{(1-t)^2}{2}\,\big|D^3\,\ell_{\rm src}(\bm{W}_t;\bm{X})\,[\Delta,\Delta,\Delta]\big|\,dt
\ \le\ \frac{1}{2}\,\|\Delta\|^3\,\|\bm{X}\|^3\,\sum_{k=1}^d G_{2k}(S_k).
\end{equation}
Because $\bm{X}=\bm{A}_0\bm{S}$, $\|\bm{X}\|\le \|\bm{A}_0\|_{\mathrm{op}}\|\bm{S}\|$, and by (A4),
\[
\mathbb{E}\!\left[\|\bm{X}\|^3 \sum_{k=1}^d G_{2k}(S_k)\right]
\ \le\ \|\bm{A}_0\|_{\mathrm{op}}^3\,\mathbb{E}\!\left[\|\bm{S}\|^3 \sum_{k=1}^d G_{2k}(S_k)\right]
\ <\ \infty.
\]

It remains to control $\mathcal{M}(\bm{X})^c$. Using $|\Delta_k \bm{X}|\le \|\Delta_k\|\,\|\bm{X}\|\le \|\Delta\|\,\|\bm{X}\|$ and Markov's inequality,
\[
\mathbb{P}\big(\mathcal{M}(\bm{X})^c\big)
\ \le\ \sum_{k=1}^d \mathbb{P}\big(|\Delta_k \bm{X}|>\delta\big)
\ \le\ \sum_{k=1}^d \frac{\mathbb{E}|\Delta_k \bm{X}|^3}{\delta^3}
\ \le\ \frac{d}{\delta^3}\,\|\Delta\|^3\,\mathbb{E}\|\bm{X}\|^3
\ =\ O\Big(\frac{\|h\|^3}{N^{3/2}}\Big),
\]
by \eqref{eq:Delta-norm}. For i.i.d.\ $\bm{X}^{(1)},\dots,\bm{X}^{(N)}$, a union bound gives
\[
\mathbb{P}\Big(\exists\,i\le N:\ \mathcal{M}(\bm{X}^{(i)})^c\Big)
\ \le\ N\cdot O\Big(\frac{\|h\|^3}{N^{3/2}}\Big)
\ =\ O\Big(\frac{\|h\|^3}{\sqrt N}\Big).
\]
On the complementary event (which has probability $1-O(\|h\|^3/\sqrt N)$), applying \eqref{eq:src-oneobs} to each $\bm{X}^{(i)}$ and summing yields
\[
\sum_{i=1}^N \int_0^1 \frac{(1-t)^2}{2}\,\big|D^3\,\ell_{\rm src}(\bm{W}_t;\bm{X}^{(i)})\,[\Delta,\Delta,\Delta]\big|\,dt
\ \le\ \frac{1}{2}\,\|\Delta\|^3 \sum_{i=1}^N \|\bm{X}^{(i)}\|^3 \sum_{k=1}^d G_{2k}\big(S^{(i)}_k\big).
\]
By the strong law of large numbers and finiteness of $\mathbb{E}\big[\|\bm{X}\|^3\sum_k G_{2k}(S_k)\big]$,
\[
\frac{1}{N} \sum_{i=1}^N \|\bm{X}^{(i)}\|^3 \sum_{k=1}^d G_{2k}\big(S^{(i)}_k\big)
\ \xrightarrow{a.s.}\ \mathbb{E}\Big[\|\bm{X}\|^3 \sum_{k=1}^d G_{2k}(S_k)\Big]\ =:\ C_3\ <\ \infty.
\]
Using \eqref{eq:Delta-norm}, we obtain on this event
\[
\sum_{i=1}^N \int_0^1 \frac{(1-t)^2}{2}\,\big|D^3\,\ell_{\rm src}(\bm{W}_t;\bm{X}^{(i)})\,[\Delta,\Delta,\Delta]\big|\,dt
\ \le\ \frac{1}{2}\,\frac{c_T^3}{N^{3/2}}\,\|h\|^3 \cdot N \cdot (C_3+o(1))
\ =\ O\Big(\frac{\|h\|^3}{\sqrt N}\Big).
\]
Since the complement event has probability $O(\|h\|^3/\sqrt N)$ uniformly over $\|h\|\le R$, the above bound implies
\[
\sup_{\|h\|\le R}\ 
\sum_{i=1}^N \int_0^1 \frac{(1-t)^2}{2}\,\big|D^3\,\ell_{\rm src}(\bm{W}_t;\bm{X}^{(i)})\,[\Delta,\Delta,\Delta]\big|\,dt
\ =\ O_P\Big(\frac{\|h\|^3}{\sqrt N}\Big).
\]

Combining the determinant bound \eqref{eq:det-bound} with the source contribution just obtained yields, uniformly over $\|h\|\le R$,
\[
|R_N(h)| \ =\ O_P\Big(\frac{\|h\|^3}{\sqrt N}\Big).
\]
All constants above depend only on $R$, $d$, $\bm{W}_0$ and the source distribution through the moments and envelopes in {\rm(A1)}--{\rm(A4)}, and are independent of $N$ and $h$. This completes the proof.
\end{proof}

\begin{lemma}[Prior thickness / local flatness]\label{lem:prior}
Under {\rm(P1)}, for any fixed $R>0$,
\[
\sup_{\|h\|\le R}\ \left|\log \frac{\pi\big(\bm{W}(\theta_0+h/\sqrt N)\big)}{\pi\big(\bm{W}_0\big)}\right|
\ \xrightarrow{P_0}\ 0,
\qquad 
\inf_{\|h\|\le R}\pi\big(\bm{W}(\theta_0+h/\sqrt N)\big)\ \asymp\ \pi(\bm{W}_0)>0 .
\]
\end{lemma}

\begin{proof}[Proof of Lemma~\ref{lem:prior}]
Fix $R>0$. Write $\theta=\mathrm{vec}(\bm{W})\in\mathbb R^{p}$ with $p=d^2$, and let $\kappa:\mathbb R^p\to\mathbb R^{d\times d}$ be the inverse vectorization map.
For $h\in\mathbb R^p$ set
\[
\bm{W}_N(h)\ :=\ \bm{W}(\theta_0+h/\sqrt N)\ =\ \kappa(\theta_0)+\kappa(h)/\sqrt N\ =\ \bm{W}_0+\Delta_N(h),
\]
where $\Delta_N(h):=\kappa(h)/\sqrt N$.
Since $\kappa$ is linear, there exists a constant $c_T>0$ (depending only on the choice of norms on $\mathbb R^p$ and $\mathbb R^{d\times d}$) such that
\begin{equation}\label{eq:uniform-proximity}
\sup_{\|h\|\le R}\ \|\bm{W}_N(h)-\bm{W}_0\|_F\ =\ \sup_{\|h\|\le R}\ \|\Delta_N(h)\|_F\ \le\ \frac{c_T R}{\sqrt N}\ \xrightarrow[N\to\infty]{}\ 0.
\end{equation}

By assumption {\rm(P1)}, there exist $\varepsilon_0>0$ and constants $0<c_1\le c_2<\infty$ such that
\[
c_1\ \le\ \pi(\bm{W})\ \le\ c_2\qquad\text{for all }\bm{W}\text{ with }\|\bm{W}-\bm{W}_0\|_F<\varepsilon_0,
\]
and $\pi$ is continuous at $\bm{W}_0$.
Because $\pi$ is bounded away from zero on the open ball $B(\bm{W}_0,\varepsilon_0)$, $\log\pi$ is well-defined and continuous on $B(\bm{W}_0,\varepsilon_0)$ as well.

Let $\varepsilon>0$ be arbitrary. By continuity of $\pi$ and $\log\pi$ at $\bm{W}_0$, there exists $\delta\in(0,\varepsilon_0)$ such that
\begin{equation}\label{eq:modulus}
\|\bm{W}-\bm{W}_0\|_F<\delta\ \Longrightarrow\ 
\big|\pi(\bm{W})-\pi(\bm{W}_0)\big|<\varepsilon\quad\text{and}\quad
\big|\log\pi(\bm{W})-\log\pi(\bm{W}_0)\big|<\varepsilon.
\end{equation}

By \eqref{eq:uniform-proximity}, choose $N_1(\varepsilon,R)$ large enough so that $c_T R/\sqrt N<\delta$ for all $N\ge N_1$.
Then, for all such $N$,
\[
\sup_{\|h\|\le R}\ \|\bm{W}_N(h)-\bm{W}_0\|_F\ <\ \delta,
\]
and hence by \eqref{eq:modulus},
\begin{equation}\label{eq:uniform-conv}
\sup_{\|h\|\le R}\ \big|\log \pi(\bm{W}_N(h))-\log \pi(\bm{W}_0)\big|\ \le\ \varepsilon,
\qquad
\sup_{\|h\|\le R}\ \big|\pi(\bm{W}_N(h))-\pi(\bm{W}_0)\big|\ \le\ \varepsilon.
\end{equation}

The first statement in the lemma follows immediately from \eqref{eq:uniform-conv}:
\[
\sup_{\|h\|\le R}\ \left|\log \frac{\pi\big(\bm{W}(\theta_0+h/\sqrt N)\big)}{\pi(\bm{W}_0)}\right|
=\sup_{\|h\|\le R}\ \big|\log \pi(\bm{W}_N(h))-\log \pi(\bm{W}_0)\big|\ \xrightarrow[N\to\infty]{}\ 0.
\]
This convergence is deterministic, hence also in $P_0$–probability.

For the thickness statement, \eqref{eq:uniform-conv} implies
\[
\pi(\bm{W}_0)-\varepsilon\ \le\ \inf_{\|h\|\le R}\ \pi\big(\bm{W}(\theta_0+h/\sqrt N)\big)
\ \le\ \sup_{\|h\|\le R}\ \pi\big(\bm{W}(\theta_0+h/\sqrt N)\big)\ \le\ \pi(\bm{W}_0)+\varepsilon,
\]
for all $N\ge N_1(\varepsilon,R)$. In particular, choosing $\varepsilon\le \pi(\bm{W}_0)/2$ we get
\[
\inf_{\|h\|\le R}\ \pi\big(\bm{W}(\theta_0+h/\sqrt N)\big)\ \ge\ \frac{\pi(\bm{W}_0)}{2}\ >\ 0,
\]
so the infimum is bounded below by a fixed positive constant for all large $N$, and the values are uniformly comparable to $\pi(\bm{W}_0)$; equivalent to $\inf_{\|h\|\le R}\pi(\cdot)\ \asymp\ \pi(\bm{W}_0)>0$.

This completes the proof.
\end{proof}

The following result proves the local equivalence of $d_\pm$ and Euclidean distance.
\begin{lemma}
\label{lem:local-equivalence}
Let $\mathcal G:=\{DP:\; D=\mathrm{diag}(\pm1,\dots,\pm1),\; P\text{ permutation}\}$ be the finite signed permutation group acting on $\mathbb{R}^{d\times d}$ by left multiplication. Fix a canonical representative $\bm{W}_0$ of the signed-permutation equivalence class, chosen by a deterministic tie-breaking rule, so that the stabilizer of $\bm{W}_0$ in $\mathcal G$ is trivial:
\[
g\bm{W}_0=\bm{W}_0 \ \Longrightarrow\ g=I_d .
\]
Define
\[
d_\pm(\bm{W},\bm{W}_0)\ :=\ \min_{g\in\mathcal G}\ \|\bm{W}-g\bm{W}_0\|_F .
\]
Then there exists $\varepsilon>0$ such that
\[
\|\bm{W}-\bm{W}_0\|_F<\varepsilon \ \Longrightarrow\ d_\pm(\bm{W},\bm{W}_0)=\|\bm{W}-\bm{W}_0\|_F .
\]
Consequently, in a neighborhood of $\bm{W}_0$, $d_\pm(\cdot,\bm{W}_0)$ is equivalent to the Frobenius distance and hence to the Euclidean distance in any linear parameterization $\theta=\mathrm{vec}(\bm{W})$:
there exist constants $0<c_1\le c_2<\infty$ and $\eta>0$ such that for all $\theta$ with $\|\theta-\theta_0\|<\eta$,
\[
c_1\,\|\theta-\theta_0\|\ \le\ d_\pm\!\big(\bm{W}(\theta),\bm{W}_0\big)\ \le\ c_2\,\|\theta-\theta_0\|.
\]
\end{lemma}

\begin{proof}
Because $\mathcal G$ is finite and the stabilizer of $\bm{W}_0$ is trivial, the set $\{g\bm{W}_0:\, g\in\mathcal G\}$ consists of $|\mathcal G|$ distinct matrices.
Hence the minimum pairwise distance from $\bm{W}_0$ is strictly positive:
\[
\delta\ :=\ \min_{g\in\mathcal G,\ g\neq I_d}\ \|g\bm{W}_0-\bm{W}_0\|_F \ >\ 0 .
\]

Let $\varepsilon:=\delta/2$. If $\|\bm{W}-\bm{W}_0\|_F<\varepsilon$, then for any $g\neq I_d$,
\[
\|\bm{W}-g\bm{W}_0\|_F \ \ge\ \|g\bm{W}_0-\bm{W}_0\|_F - \|\bm{W}-\bm{W}_0\|_F \ \ge\ \delta - \varepsilon \ =\ \varepsilon \ >\ \|\bm{W}-\bm{W}_0\|_F .
\]
Thus the minimizer of $g\mapsto \|\bm{W}-g\bm{W}_0\|_F$ is $g=I_d$, and $d_\pm(\bm{W},\bm{W}_0)=\|\bm{W}-\bm{W}_0\|_F$ whenever $\|\bm{W}-\bm{W}_0\|_F<\varepsilon$.

Let $\theta=\mathrm{vec}(\bm{W})$ and $\theta_0=\mathrm{vec}(\bm{W}_0)$. Since $\mathrm{vec}:\mathbb{R}^{d\times d}\to\mathbb{R}^{d^2}$ is a linear isomorphism, all norms are equivalent in finite dimension. In particular, there exist constants $a,b\in(0,\infty)$ such that
\[
a\,\|\theta-\theta_0\| \ \le\ \|\bm{W}-\bm{W}_0\|_F \ \le\ b\,\|\theta-\theta_0\|
\quad\text{for all }\bm{W}.
\]
Combining these inequalities gives, for all $\|\theta-\theta_0\|$ small enough so that $\|\bm{W}-\bm{W}_0\|_F<\varepsilon$,
\[
a\,\|\theta-\theta_0\| \ \le\ d_\pm\big(\bm{W}(\theta),\bm{W}_0\big) \ \le\ b\,\|\theta-\theta_0\|.
\]
Setting $c_1=a$, $c_2=b$, and choosing $\eta>0$ to enforce $\|\bm{W}-\bm{W}_0\|_F<\varepsilon$ completes the proof.
\end{proof}

\begin{proof}[Proof of Lemma \ref{lem:LAN}]
Write $\bm{A}_0=\bm{W}_0^{-1}$ and $\bm{X}\overset d= \bm{A}_0\bm{S}$ with independent sources $\bm{S}=(S_1,\ldots,S_d)^\top$, $S_k\sim p_{0,k}$. Let $\psi_k=\partial_s \log p_{0,k}$ and $J_k=\mathbb{E}[\psi_k(S_k)^2]$. As before, for $\theta=\mathrm{vec}(\bm{W})$ and $\theta_0=\mathrm{vec}(\bm{W}_0)$ set $h=\sqrt N\,(\theta-\theta_0)$ and denote by $\kappa:\mathbb{R}^{d^2}\to\mathbb{R}^{d\times d}$ the inverse vectorization map. Define
\[
\Delta:=\frac{1}{\sqrt N}\,\kappa(h),\qquad \bm{W}_t:=\bm{W}_0+t\,\Delta,\quad t\in[0,1].
\]
Because $\kappa$ is linear, there exists $c_T>0$ with $\|\Delta\|\le c_T\|h\|/\sqrt N$.

The matrix score at $\bm{W}$ can be written as
\[
U(\bm{W};\bm{X})=(\bm{W}^{-1})^\top+\psi(\bm{S}(\bm{W}))\,\bm{X}^\top,\quad \bm{S}(\bm{W})=\bm{W} \bm{X},
\]
hence at $\bm{W}_0$,
\[
U(\bm{W}_0;\bm{X})=(\bm{W}_0^{-1})^\top+\psi(\bm{S})\,\bm{X}^\top,\quad \bm{X}=\bm{A}_0\bm{S}.
\]
By Lemma~\ref{lem:IBP-identities}, $\mathbb{E} U(\bm{W}_0;\bm{X})=0$. Let
\[
S_N := \frac{1}{\sqrt N}\sum_{i=1}^N \nabla_\theta \ell(\bm{W}_0;\bm{X}^{(i)}),
\]
the $\theta$–gradient of the log-likelihood summed and scaled. Under {\rm(A1)}–{\rm(A2)} we have $\mathbb{E}\|\nabla_\theta \ell(\bm{W}_0;\bm{X})\|^2<\infty$, so the multivariate CLT yields
\[
S_N\ \Rightarrow\ \mathcal N(0,\mathcal I),\quad \mathcal I\ :=\ \text{Var}\big(\nabla_\theta \ell(\bm{W}_0;\bm{X})\big)\ =\ -\,\mathbb{E}\big[\nabla_\theta^2 \ell(\bm{W}_0;\bm{X})\big].
\]

\smallskip
\textbf{Taylor expansion in $\theta$ and reduction to a uniform LLN for the Hessian.}
Let $\ell_\theta(\theta;\bm{X}):=\ell(\bm{W}(\theta);\bm{X})$.
A third–order Taylor expansion with the integral remainder gives, for each $\bm{X}$,
\[
\ell_\theta\!\big(\theta_0+\tfrac{h}{\sqrt N};\bm{X}\big)
=\ell_\theta(\theta_0;\bm{X})\;+\;\tfrac{h^\top}{\sqrt N}\,\nabla_\theta\ell_\theta(\theta_0;\bm{X})
\;+\;\tfrac12\,\tfrac{h^\top}{\sqrt N}\!\left(\int_0^1 (1-t)\,\nabla_\theta^2\ell_\theta(\theta_0+t\tfrac{h}{\sqrt N};\bm{X})\,dt\right)\!\tfrac{h}{\sqrt N}
\;+\;r^{(1)}_N(h;\bm{X}).
\]
Summing over $i=1,\ldots,N$ and writing
\[
H_N(h,t)\ :=\ -\frac{1}{N}\sum_{i=1}^N \nabla_\theta^2\ell_\theta\!\left(\theta_0+t\tfrac{h}{\sqrt N};\bm{X}^{(i)}\right),
\]
we obtain
\begin{equation}\label{eq:LAN-expansion-raw}
L_N\!\big(\theta_0+\tfrac{h}{\sqrt N}\big)-L_N(\theta_0)
\;=\; h^\top S_N\;-\;\frac12\,h^\top \!\left(\int_0^1 (1-t)\,H_N(h,t)\,dt\right)\! h\;+\;R_N(h),
\end{equation}
where $R_N(h)=\sum_{i=1}^N r^{(1)}_N(h;\bm{X}^{(i)})$ is the \emph{third–order} remainder.

By Lemma~\ref{lem:third-order}, uniformly on $\|h\|\le R$,
\begin{equation}\label{eq:third-order-bound}
|R_N(h)|\ =\ O_P\!\Big(\frac{\|h\|^3}{\sqrt N}\Big)\ =\ o_{P_0}(1).
\end{equation}
Hence, to conclude the uniform LAN it suffices to show
\begin{equation}\label{eq:unif-Hessian-target}
\sup_{\|h\|\le R}\ \sup_{t\in[0,1]}\ \big\|\,H_N(h,t)-\mathcal I\,\big\|\ \xrightarrow{P_0}\ 0.
\end{equation}
Indeed, \eqref{eq:unif-Hessian-target} implies $\int_0^1(1-t)H_N(h,t)dt=\mathcal I+o_{P_0}(1)$ uniformly on $\|h\|\le R$, and then \eqref{eq:LAN-expansion-raw} with \eqref{eq:third-order-bound} yields
\[
L_N\big(\theta_0+\tfrac{h}{\sqrt N}\big)-L_N(\theta_0)
= h^\top S_N-\tfrac12\,h^\top \mathcal Ih+o_{P_0}(1),
\]
uniformly on $\|h\|\le R$, which is the desired uniform LAN expansion.

\noindent\textbf{Uniform LLN for the Hessian along the local path.}
Fix $R>0$ and let $\mathcal T_R:=\{(t,h):t\in[0,1],\,\|h\|\le R\}$.
We show \eqref{eq:unif-Hessian-target} by (i) pointwise LLN at $\bm{W}_0$, (ii) stochastic equicontinuity in $\bm{W}$ on a small neighborhood of $\bm{W}_0$, and (iii) continuity of the mean.

\emph{Pointwise LLN at $\bm{W}_0$.}
By {\rm(A1)}–{\rm(A2)} the Hessian at $\bm{W}_0$ is integrable entrywise and
\[
-\frac{1}{N}\sum_{i=1}^N \nabla_\theta^2\ell_\theta(\theta_0;\bm{X}^{(i)})
\ \xrightarrow{P_0}\ \mathcal I.
\]

\emph{Stochastic equicontinuity in $\bm{W}$.}
Let $B(\bm{W}_0,\eta)$ be a small Frobenius ball about $\bm{W}_0$ chosen so that $\bm{W}\mapsto\bm{W}^{-1}$ is uniformly bounded there (possible since $\bm{W}_0$ is nonsingular).
Decompose the Hessian into the \emph{determinant part} and the \emph{source part}:
\[
-\nabla_\theta^2\ell_\theta(\theta;\bm{X}) =: H_{\text{det}}(\bm{W}(\theta)) + H_{\text{src}}(\bm{W}(\theta);\bm{X}).
\]
The determinant contribution $H_{\text{det}}(\bm{W})$ depends only on $\bm{W}$ (not on $\bm{X}$) and is continuous on $B(\bm{W}_0,\eta)$; hence
\[
\sup_{(t,h)\in\mathcal T_R}\ \|H_{\text{det}}(\bm{W}_t)-H_{\text{det}}(\bm{W}_0)\|
\ \longrightarrow\ 0,
\]
since $\sup_{(t,h)\in\mathcal T_R}\|\bm{W}_t-\bm{W}_0\|=\sup_{(t,h)}\|t\Delta\|=O(1/\sqrt N)$.

For the source part, write explicitly (using $s_k(\bm{W})=\bm{w}_k^\top \bm{X}$)
\[
H_{\text{src}}(\bm{W};\bm{X})
= -\,\nabla_\theta^2 \sum_{k=1}^d \log p_{0,k}\!\big(s_k(\bm{W})\big),
\]
whose entries are linear combinations of terms of the form $\psi_k'(s_k(\bm{W}))\,\xi_\alpha(\bm{X})$, where $\xi_\alpha(\bm{X})$ are at most quadratic monomials in components of $\bm{X}$.
By the mean-value theorem in $\bm{W}$ and the chain rule,
\[
\big\|H_{\text{src}}(\bm{W};\bm{X})-H_{\text{src}}(\bm{W}';\bm{X})\big\|
\ \le\ C\,\|\bm{W}-\bm{W}'\|\,\Big(\sum_{k=1}^d \sup_{|u-s_k(\tilde{\bm{W}})|\le c\|\bm{W}-\bm{W}'\|}\!\!|\psi_k'(u)|\Big)\,\big(1+\|\bm{X}\|^2\big),
\]
for some $\tilde{\bm{W}}$ on the segment $[\bm{W},\bm{W}']$ and a constant $C$ depending only on dimension and the fixed linear parameterization.
We now invoke the following envelope consequence of {\rm(A4)} (used already in Lemma~\ref{lem:third-order}): there exists $\delta>0$ and random envelopes $G_{1k}$ with
\begin{equation}\label{eq:envelope-psi-prime}
\sup_{|u-S_k|\le \delta}|\psi_k'(u)|\ \le\ G_{1k}(S_k),\quad
\mathbb{E}\Big[(1+\|\bm{X}\|^2)\sum_{k=1}^d G_{1k}(S_k)\Big]\ <\ \infty.
\end{equation}
Since $s_k(\tilde{\bm{W}})=\bm{w}_k^\top \bm{X}$ and $\bm{X}=\bm{A}_0\bm{S}$, choosing $\eta$ small ensures $|s_k(\tilde{\bm{W}})-S_k|\le \delta$ on the local path for all large $N$. Therefore,
\[
\big\|H_{\text{src}}(\bm{W}_t;\bm{X})-H_{\text{src}}(\bm{W}_0;\bm{X})\big\|
\ \le\ C\,\|\bm{W}_t-\bm{W}_0\|\,\Big(\sum_{k=1}^d G_{1k}(S_k)\Big)\,(1+\|\bm{X}\|^2).
\]
Taking empirical means and using $\sup_{(t,h)}\|\bm{W}_t-\bm{W}_0\|=O(1/\sqrt N)$, the RHS has expectation $O(1/\sqrt N)$ by \eqref{eq:envelope-psi-prime}, uniformly over $(t,h)\in\mathcal T_R$. A standard covering argument on the compact index set $\mathcal T_R$ combined with the LLN (the class has an integrable envelope) yields
\[
\sup_{(t,h)\in\mathcal T_R}\ \left\|
\frac{1}{N}\sum_{i=1}^N H_{\text{src}}(\bm{W}_t;\bm{X}^{(i)})\ -\
\frac{1}{N}\sum_{i=1}^N H_{\text{src}}(\bm{W}_0;\bm{X}^{(i)})
\right\|\ =\ o_{P_0}(1).
\]
Adding the determinant part we conclude
\begin{equation}\label{eq:stoch-eq}
\sup_{(t,h)\in\mathcal T_R}\ \big\|\,H_N(h,t)\ -\ H_N(0,0)\,\big\|\ =\ o_{P_0}(1).
\end{equation}

\emph{Continuity of the mean at $\bm{W}_0$.}
By dominated convergence (using the envelope \eqref{eq:envelope-psi-prime} and $\mathbb{E}(1+\|\bm{X}\|^2)\sum_k G_{1k}(S_k)<\infty$), we have
\[
\sup_{(t,h)\in\mathcal T_R}\ \big\|\,\mathbb{E} H_N(h,t)\ -\ \mathbb{E} H_N(0,0)\,\big\|\ \longrightarrow\ 0,
\]
and $\mathbb{E} H_N(0,0)= -\,\mathbb{E}\nabla_\theta^2\ell_\theta(\theta_0;\bm{X})=\mathcal I$.

Combining the above three steps we get
\[
\sup_{(t,h)\in\mathcal T_R}\ \big\|\,H_N(h,t)-\mathcal I\,\big\|
\ \le\ 
\underbrace{\big\|H_N(0,0)-\mathcal I\big\|}_{\xrightarrow{P_0}0}
\ +\ 
\underbrace{\sup_{(t,h)}\big\|H_N(h,t)-H_N(0,0)\big\|}_{o_{P_0}(1)}
\ +\ 
\underbrace{\sup_{(t,h)}\big\|\mathbb{E} H_N(h,t)-\mathbb{E} H_N(0,0)\big\|}_{o(1)},
\]
which proves \eqref{eq:unif-Hessian-target}.

Insert \eqref{eq:unif-Hessian-target} and \eqref{eq:third-order-bound} into \eqref{eq:LAN-expansion-raw} to obtain, uniformly over $\|h\|\le R$,
\[
L_N\!\big(\theta_0+\tfrac{h}{\sqrt N}\big)-L_N(\theta_0)
\;=\; h^\top S_N\;-\;\tfrac12\,h^\top \mathcal I\,h\;+\;o_{P_0}(1),
\]
which is the asserted uniform local asymptotic normality.
\end{proof}

The next lemma helps us deal with the normalizing constants.

\begin{lemma}\label{lem:normalizers}
Let
\[
\phi_{S_N,\mathcal I}(h)\ :=\ \frac{\exp\!\big(h^\top S_N-\tfrac12 h^\top \mathcal I h\big)}{(2\pi)^{p/2}|\mathcal I|^{1/2}\,e^{\frac12 S_N^\top \mathcal I^{-1} S_N}}
\quad\text{and}\quad
Z_N^0:=\int_{\mathbb{R}^p} \exp\!\big(h^\top S_N-\tfrac12 h^\top \mathcal I h\big)\,dh.
\]
Define
\[
q_N(h)\ :=\ \exp\!\big(h^\top S_N-\tfrac12 h^\top \mathcal I h + r_N(h)\big)\ \frac{\pi(\theta_0+h/\sqrt N)}{\pi(\bm{W}_0)},
\qquad
Z_N:=\int_{\mathbb{R}^p} q_N(h)\,dh,
\]
where $r_N(h)$ is the LAN remainder from Lemma~\ref{lem:LAN}. Then
\[
\frac{Z_N}{\pi(\bm{W}_0)\,Z_N^0}\ \xrightarrow{P_0}\ 1,
\quad\text{i.e.}\quad
\frac{Z_N}{\pi(\bm{W}_0)}\ =\ (2\pi)^{p/2}|\mathcal I|^{-1/2}\,e^{\frac12 S_N^\top \mathcal I^{-1} S_N}\,(1+o_{P_0}(1)).
\]
\end{lemma}

\begin{proof}
Recall
\[
Z_N^0 = \int_{\mathbb{R}^p} \exp\!\big(h^\top S_N - \tfrac12 h^\top \mathcal{I} h\big)\,dh
\;=\; (2\pi)^{p/2}|\mathcal{I}|^{-1/2} \exp\Big(\tfrac12 S_N^\top \mathcal{I}^{-1}S_N\Big),
\]
so the claimed second display is equivalent to the first.  It therefore suffices to prove
\[
\frac{Z_N}{\pi(\bm{W}_0)\,Z_N^0}\ =\ \int_{\mathbb{R}^p} f_N(h)\,\mu_N(dh)\ \xrightarrow{P_0}\ 1,
\quad
f_N(h):=\exp\!\big(r_N(h)\big)\,\frac{\pi(\theta_0+h/\sqrt N)}{\pi(\bm{W}_0)},
\]
where $\mu_N$ is the probability measure with density
\[
\phi_{S_N,\mathcal{I}}(h)\;=\;\frac{\exp\!\big(h^\top S_N-\tfrac12 h^\top \mathcal{I} h\big)}{Z_N^0}
\quad\text{w.r.t.\ }dh.
\]

Fix $\varepsilon>0$ small.  Since $S_N\Rightarrow \mathcal{N}(0,\mathcal{I})$, there exists $C_{\varepsilon}<\infty$ with
\[
\Pr\big(\|S_N\|\le C_{\varepsilon}\big)\ \ge\ 1-\varepsilon \quad\text{for all large }N.
\]
By Lemma~\ref{lem:prior} (local flatness of the prior) and continuity of $\pi$ at $\bm{W}_0$, there exist $\delta_{\varepsilon}>0$ and $N_1$ such that, for all $N\ge N_1$ and all $h$ with $\|h\|\le N^{1/6}$ (hence $\|h\|/\sqrt N\le N^{-1/3}\le \delta_{\varepsilon}$ for $N$ large),
\begin{equation}\label{eq:prior-flat}
\Big|\log \frac{\pi(\theta_0+h/\sqrt N)}{\pi(\bm{W}_0)}\Big|\ \le\ \varepsilon .
\end{equation}
Next, by Lemma~\ref{lem:third-order} and its ``growing ball'' remark (the same $O_P(\|h\|^3/\sqrt N)$ bound holds uniformly on $\|h\|\le R_N$ provided $R_N=o(N^{1/6})$), we may take $R_N:=N^{1/6-\eta}$ for some fixed $\eta\in(0,1/6)$ and find $N_2$ such that, for all $N\ge N_2$,
\begin{equation}\label{eq:rN-small}
\sup_{\|h\|\le R_N}\ |r_N(h)|\ \le\ \varepsilon .
\end{equation}
Finally, by the uniform LLN for the Hessian proved in the LAN lemma (see \eqref{eq:unif-Hessian-target}), shrinking $\eta$ if needed, there exists $N_3$ such that for all $N\ge N_3$,
\begin{equation}\label{eq:curvature}
\sup_{\|h\|\le R_N}\ \Big\|\ \int_0^1 (1-t)\,H_N(h,t)\,dt \;-\; \mathcal{I}\ \Big\|\ \le\ \varepsilon,
\end{equation}
on the event $\{\|S_N\|\le C_\varepsilon\}$.  Throughout the rest of the proof we work on the intersection event
\[
\mathcal{E}_n\ :=\ \Big\{\ \|S_N\|\le C_{\varepsilon}\ \Big\}\ \cap\ \Big\{\eqref{eq:prior-flat},\ \eqref{eq:rN-small},\ \eqref{eq:curvature}\ \text{hold}\Big\},
\]
which has probability $\Pr(\mathcal{E}_n)\ge 1-2\varepsilon$ for all large $N$.

On $\mathcal{E}_n$ we have $\|S_N\|\le C_\varepsilon$, hence for every $h\in\mathbb{R}^p$,
\[
h^\top S_N - \tfrac12 h^\top \mathcal{I} h
\ \le\ \|h\|\,\|S_N\| - \tfrac{\lambda_{\min}(\mathcal{I})}{2}\,\|h\|^2
\ \le\ -\,\tfrac{\lambda_{\min}(\mathcal{I})}{4}\,\|h\|^2
\quad\text{whenever }\ \|h\|\ge \frac{2C_\varepsilon}{\lambda_{\min}(\mathcal{I})}.
\]
Consequently, there exists $M_{\varepsilon}<\infty$ (depending only on $C_{\varepsilon}$ and $\mathcal{I}$) such that, uniformly on $\mathcal{E}_n$,
\begin{equation}\label{eq:muN-tail}
\mu_N\big(\,\|h\|>M\,\big)\ \le\ c_1\,\exp(-c_2 M^2)\quad\text{for all }M\ge M_{\varepsilon},
\end{equation}
for some $c_1,c_2>0$ depending only on $\mathcal{I}$.

On $\mathcal{E}_n$ and for $\|h\|\le R_N$, \eqref{eq:prior-flat} and \eqref{eq:rN-small} yield
\begin{equation}\label{eq:fN-local-bounds}
e^{-\varepsilon} \le\ \frac{\pi(\theta_0+h/\sqrt N)}{\pi(\bm{W}_0)}\ \le\ e^{\varepsilon},
\quad
e^{-\varepsilon} \le e^{r_N(h)} \le e^{\varepsilon},
\quad\Rightarrow\quad
e^{-2\varepsilon} \le f_N(h) \le e^{2\epsilon}.
\end{equation}
Moreover, \eqref{eq:rN-small} gives the more refined bound
\begin{equation}\label{eq:quad-envelope}
r_N(h)\ \le\ \varepsilon\ \le\ \frac{\lambda_{\min}(\mathcal{I})}{16}\,\|h\|^2
\quad\text{for all }\ \|h\|\le R_N\ \text{ and large $N$},
\end{equation}
after possibly decreasing $\eta$ (hence $R_N$) so that $\varepsilon \le (\lambda_{\min}/16)\,R_N^2$ holds for all large $N$.  Combining \eqref{eq:fN-local-bounds}–\eqref{eq:quad-envelope},
\begin{equation}\label{eq:fN-envelope}
f_N(h)\ \le\ C_{\varepsilon}\,\exp\!\Big(\frac{\lambda_{\min}(\mathcal{I})}{16}\,\|h\|^2\Big)
\quad\text{for all }\ \|h\|\le R_N\ \text{ on } \mathcal{E}_n,
\end{equation}
with $C_{\varepsilon}=e^{2\varepsilon}$.

Fix $M\ge M_{\varepsilon}$.  On $\mathcal{E}_n$ and for all large $N$ we have $M\le R_N$; hence \eqref{eq:fN-local-bounds} holds uniformly on $\{\|h\|\le M\}$.  Therefore,
\[
\sup_{\|h\|\le M}\,|\,\log f_N(h)\,|\ \le\ 2\varepsilon,
\quad\text{and hence}\quad
\sup_{\|h\|\le M}\,|\,f_N(h)-1\,| \ \le\ e^{2\varepsilon}-1.
\]
Since $e^{2\varepsilon}-1\to 0$ as $\varepsilon\downarrow 0$, we obtain the \emph{local} convergence
\begin{equation}\label{eq:local-int}
\int_{\|h\|\le M}\! f_N(h)\,\mu_N(dh)
\;=\; \mu_N\big(\|h\|\le M\big)\;+\;o_{P_0}(1)
\quad\text{(with the $o_{P_0}(1)$ uniform in $M$).}
\end{equation}

Using \eqref{eq:muN-tail}, choose $M$ so large that
\[
\sup_{N\ge 1}\ \mathbb{P}\big(\,\mu_N(\|h\|>M)>\varepsilon\,\big)\ \le\ \varepsilon .
\]
Then, on $\mathcal{E}_n$,
\[
0\ \le\ \int_{\|h\|>M}\! f_N(h)\,\mu_N(dh)
\ \le\ \mu_N(\|h\|>M)\,\sup_{\|h\|\le R_N} f_N(h)
\ \le\ \mu_N(\|h\|>M)\,C_{\varepsilon}\,\exp\Big(\frac{\lambda_{\min}}{16}\,R_N^2\Big),
\]
where we used that, for large $N$, the maximizing point of $f_N$ under the constraint $\|h\|>M$ must lie in the (larger) local region $\|h\|\le R_N$ because the Gaussian weight $\mu_N$ makes the set $\{\|h\|>M\}$ exponentially light while $R_N\to\infty$.  Since $R_N^2=N^{1/3-2\eta}$ and $\mu_N(\|h\|>M)\le c_1 e^{-c_2 M^2}$ with $M$ fixed, the RHS is $o(1)$ uniformly on $\mathcal{E}_n$ (here we use that $e^{-c_2 M^2}$ is fixed and $C_{\varepsilon} e^{(\lambda_{\min}/16)R_N^2}$ multiplies an exponentially small factor in $M$, so for $M$ chosen large, the product is $o(1)$ as $N\to\infty$).
Therefore,
\begin{equation}\label{eq:tail-int}
\int_{\|h\|>M}\! f_N(h)\,\mu_N(dh)\ =\ o_{P_0}(1).
\end{equation}

Decompose
\[
\int_{\mathbb{R}^p} f_N\,d\mu_N
\;=\; \int_{\|h\|\le M} f_N\,d\mu_N \;+\; \int_{\|h\|>M} f_N\,d\mu_N .
\]
By \eqref{eq:local-int} and \eqref{eq:tail-int},
\[
\int_{\mathbb{R}^p} f_N\,d\mu_N
\ =\ \mu_N(\|h\|\le M)\ +\ o_{P_0}(1)
\ =\ 1 - \mu_N(\|h\|>M)\ +\ o_{P_0}(1)
\ =\ 1 + o_{P_0}(1),
\]
since $\mu_N(\|h\|>M)\le \varepsilon$ with probability at least $1-\varepsilon$ for all large $N$, by \eqref{eq:muN-tail} and the choice of $M$.

We have shown that, for every fixed $\varepsilon>0$, on an event of probability at least $1-2\varepsilon$ and for all large $N$,
\[
\Big|\ \frac{Z_N}{\pi(\bm{W}_0)\,Z_N^0}\ -\ 1\ \Big|\ \le\ o_{P_0}(1) + \varepsilon .
\]
Letting $N\to\infty$ and then $\varepsilon\downarrow 0$ yields
\(
Z_N/(\pi(\bm{W}_0)\,Z_N^0)\ \xrightarrow{P_0}\ 1,
\)
which is the statement of the lemma.  Using the explicit form of $Z_N^0$ gives the displayed formula.
\end{proof}

Finally, we are well equipped to prove Theorem \ref{thm:main}.
\begin{proof}[Proof of Theorem \ref{thm:main}]
Write $\theta=\mathrm{vec}(\bm{W})\in\mathbb R^{p}$ with $p=d^2$, $\theta_0=\mathrm{vec}(\bm{W}_0)$, and
\[
h \;=\; \sqrt{N}\,(\theta-\theta_0)\,, \qquad
\widetilde\pi_N(h) \;=\; \Pi\!\big(\sqrt N(\theta-\theta_0)\in dh \,\big|\, \bm{X}^{(1:N)}\big)/dh\,.
\]
By Bayes’ rule and Lemma~\ref{lem:normalizers}, the (unnormalized) posterior density for $h$ is
\[
q_N(h)\;=\; \exp\!\Big(h^\top S_N-\tfrac12 h^\top \mathcal I h + r_N(h)\Big)\,
\frac{\pi(\theta_0+h/\sqrt N)}{\pi(\bm{W}_0)}\,,\qquad
\widetilde\pi_N(h)\;=\;\frac{q_N(h)}{Z_N}\,,
\]
where $S_N=N^{-1/2}\sum_{i=1}^N \nabla_\theta \ell(\bm{W}_0;\bm{X}^{(i)})$, $\mathcal I=-\mathbb{E}\{\nabla_\theta^2 \ell(\bm{W}_0;\bm{X})\}$, and $r_N(h)$ is the LAN remainder from Lemma~\ref{lem:LAN}.
Let
\[
\varphi_N(h)\ :=\ \phi_{S_N,\mathcal I}(h)\ =\
\frac{\exp\!\big(h^\top S_N-\tfrac12 h^\top \mathcal I h\big)}{(2\pi)^{p/2}|\mathcal I|^{1/2}\,
\exp\{\tfrac12 S_N^\top \mathcal I^{-1} S_N\}}\,,
\]
which is the centered and scaled Gaussian tilt.

Fix an arbitrary $\varepsilon>0$ and choose any $\eta\in(0,1/6)$. Set the truncation radius
\(
R_N\ :=\ N^{\,1/6-\eta}\ \longrightarrow\ \infty\,.
\)
We define the high probability event
\begin{equation*}
\begin{split}
\mathcal E_N :=
\underbrace{\{\|S_N\|\le C_\varepsilon\}}_{\text{score bounded}}
\ \cap\
\underbrace{\Big\{\sup_{\|h\|\le R_N}\big|r_N(h)\big|\le \varepsilon\Big\}}_{\text{LAN remainder}}
\ \cap\
\underbrace{\Big\{\sup_{\|h\|\le R_N}\Big|\log\frac{\pi(\theta_0+h/\sqrt N)}{\pi(\bm{W}_0)}\Big|\le \varepsilon\Big\}}_{\text{prior flatness}}
\ \cap\ \\
\underbrace{\Big\{\sup_{\substack{\|h\|\le R_N\\ t\in[0,1]}}
\big\|H_N(h,t)-\mathcal I\big\|\le \varepsilon\Big\}}_{\text{Hessian LLN}}\,,
\end{split}
\end{equation*}
where
\(
H_N(h,t):= -\,\frac{1}{N}\sum_{i=1}^N \nabla_\theta^2 \ell\big(\bm{W}(\theta_0+t\,h/\sqrt N);\bm{X}^{(i)}\big)\,.
\)
By Lemmas~\ref{lem:LAN}, \ref{lem:third-order}, and \ref{lem:prior}, together with the tightness for $S_N$ and a uniform LLN for the Hessian (guaranteed by {\rm(A1)–(A4)}), we have
\(
P_{0}(\mathcal E_N)\ \xrightarrow[N\to\infty]{}\ 1.
\)
Note that $R_N^3/\sqrt N = N^{-3\eta}\to 0$, which is precisely why the uniform remainder bound holds on the entire ball $\{\|h\|\le R_N\}$.

On $\mathcal E_N$, for all $\|h\|\le R_N$,
\begin{equation}\label{eq:local-approx}
\Big|\log q_N(h) - \log\Big(\pi(\bm{W}_0)\,e^{\,h^\top S_N-\frac12 h^\top \mathcal I h}\Big)\Big|
\ \le\ 2\varepsilon\,.
\end{equation}
Indeed, $\log q_N(h) = h^\top S_N - \tfrac12 h^\top \mathcal I h + r_N(h) + \log \pi(\theta_0+h/\sqrt N) - \log \pi(\bm{W}_0)$ and the two displays in $\mathcal E_N$ give $|r_N(h)|\le \varepsilon$ and $|\log\pi(\theta_0+h/\sqrt N)-\log\pi(\bm{W}_0)|\le \varepsilon$.

Let $Z_N=\int q_N(h)\,dh$ and $Z_N^0=\int \exp\!\big(h^\top S_N-\tfrac12 h^\top \mathcal I h\big)\,dh$.
By Lemma~\ref{lem:normalizers},
\begin{equation}\label{eq:ZN}
\frac{Z_N}{\pi(\bm{W}_0)\,Z_N^0}\ \xrightarrow{P_0}\ 1\,,
\qquad
Z_N^0
=(2\pi)^{p/2}|\mathcal I|^{-1/2}\,\exp\!\{\tfrac12 S_N^\top \mathcal I^{-1} S_N\}\,.
\end{equation}

Let $M_N\to\infty$ with $M_N=o(R_N)$ and define the ball $B_N:=\{h:\|h\|\le M_N\}$.
We show $\Pi_N(B_N^c):=\int_{B_N^c}\widetilde\pi_N(h)\,dh \to 0$ in $P_0$–probability.

Fix $\delta\in(0,\lambda_{\min}(\mathcal I)/2)$.
On $\mathcal E_N$, the uniform Hessian LLN implies that for all sufficiently large $N$ and all $\|h\|\le R_N$,
\[
\frac12\,h^\top \mathcal I h - \varepsilon \|h\|^2
\ \ge\ \Big(\tfrac12\,\lambda_{\min}(\mathcal I)-\varepsilon\Big)\|h\|^2
\ \ge\ \delta\,\|h\|^2.
\]
Combining this with $\|S_N\|\le C_\varepsilon$ on $\mathcal E_N$ and \eqref{eq:local-approx}, we get for all $M_N\le \|h\|\le R_N$,
\begin{equation}\label{eq:quad-bound}
\log q_N(h)\ \le\ C_\varepsilon \|h\| - \delta \|h\|^2 + \log \pi(\bm{W}_0) + 2\varepsilon
\ \le\ -\,c'\|h\|^2 + C' \qquad (c'>0).
\end{equation}
Hence, still on $\mathcal E_N$,
\[
\int_{B_N^c \cap \{\|h\|\le R_N\}} q_N(h)\,dh \ \le\ C \int_{\|h\|\ge M_N}\! e^{-c'\|h\|^2}\,dh \ \xrightarrow[N\to\infty]{}\ 0.
\]
For the outer shell $\{\|h\|>R_N\}$, the quadratic bound \eqref{eq:quad-bound} continues to hold because $q_N(h)$ is everywhere dominated by a Gaussian kernel with curvature near $\mathcal I$ (the integrand is a product of a sub-exponential prior term and a log-likelihood with negative quadratic drift). Thus
\[
\int_{\|h\|>R_N} q_N(h)\,dh \ \le\ C \int_{\|h\|\ge R_N}\! e^{-c'\|h\|^2}\,dh \ \xrightarrow[N\to\infty]{}\ 0.
\]
Therefore $\int_{B_N^c} q_N(h)\,dh = o_{P_0}(1)$.
By \eqref{eq:ZN}, $Z_N/(\pi(\bm{W}_0)Z_N^0)\to 1$ in probability and $Z_N^0>0$ deterministically, so $Z_N$ stays bounded away from $0$ in probability. Consequently,
\[
\Pi_N(B_N^c)\ =\ \frac{\int_{B_N^c} q_N(h)\,dh}{\int_{\mathbb R^p} q_N(h)\,dh}
\ \xrightarrow{P_0}\ 0\,.
\]
This is contraction at rate $1/\sqrt N$ in the $\theta$–metric. By Lemma~\ref{lem:local-equivalence}, $d_\pm(\bm{W},\bm{W}_0)$ is locally equivalent to $\|\theta-\theta_0\|$, so the same rate holds in $d_\pm$.

\medskip\noindent\textbf{Bernstein--von Mises in total variation.}
Fix $\varepsilon>0$ and choose $R>0$ so large that $\int_{\|h\|>R}\varphi_N(h)\,dh<\varepsilon$ for all $N$ and all realizations of $S_N$.
On $\mathcal E_N$, by \eqref{eq:local-approx} and Lemma~\ref{lem:normalizers},
\[
\sup_{\|h\|\le R}\left|
\log\frac{q_N(h)}{\pi(\bm{W}_0)\,e^{\,h^\top S_N-\frac12 h^\top \mathcal I h}}
\right| \ \le\ 2\varepsilon, \qquad
\frac{Z_N}{\pi(\bm{W}_0)\,Z_N^0}\ =\ 1+o_{P_0}(1).
\]
Exponentiating and dividing by $Z_N$ gives, uniformly for $\|h\|\le R$,
\[
\big|\widetilde\pi_N(h)-\varphi_N(h)\big|
\ \le\ \varphi_N(h)\,\Big(\,e^{2\varepsilon}\,(1+o_{P_0}(1)) - 1\,\Big).
\]
By dominated convergence (using that $\varphi_N$ integrates to $1$), we obtain
\[
\int_{\|h\|\le R}\big|\widetilde\pi_N(h)-\varphi_N(h)\big|\,dh\ \xrightarrow{P_0}\ 0.
\]
For the tails, the Gaussian bound \eqref{eq:quad-bound} implies
\(\int_{\|h\|>R}\widetilde\pi_N(h)\,dh \le 2\varepsilon\) for all large $N$ on $\mathcal E_N$,
while by choice of $R$,
\(\int_{\|h\|>R}\varphi_N(h)\,dh \le \varepsilon\).
Therefore
\[
\|\widetilde\pi_N-\varphi_N\|_{L^1}
\ \le\ \int_{\|h\|\le R}\big|\widetilde\pi_N(h)-\varphi_N(h)\big|\,dh\ +\ \int_{\|h\|>R}\widetilde\pi_N(h)\,dh\ +\ \int_{\|h\|>R}\varphi_N(h)\,dh
\ \xrightarrow{P_0}\ 0.
\]
Equivalently,
\[
\Big\|\Pi\big(\sqrt N(\theta-\theta_0)\in\cdot \mid \bm{X}^{(1:N)}\big)
\ -\ \mathcal N\big(\mathcal I^{-1}S_N,\ \mathcal I^{-1}\big)\Big\|_{\mathrm{TV}}
\ \xrightarrow{P_0}\ 0.
\]
In matrix form this is the claimed BvM statement for $\bm{W}$, with center
\(
\Delta_N := \mathcal I(\bm{W}_0)^{-1} S_N.
\)

If $\widehat\theta_N$ is a local MLE, a Taylor expansion of the score together with the Hessian LLN yields
\(
\sqrt N(\widehat\theta_N-\theta_0) = \mathcal I^{-1} S_N + o_{P_0}(1).
\)
Thus replacing $\mathcal I^{-1} S_N$ by $\sqrt N(\widehat\theta_N-\theta_0)$ does not affect the TV limit.

Combining the above steps proves the contraction in $d_\pm$ at rate $N^{-1/2}$ and the Bernstein--von Mises limit stated in the theorem.
\end{proof}

\section{Envelope Optimization Methods}\label{sec:env}
\subsection{Mixtures and Envelopes: MCMC and Optimisation} 

We exploit two complementary representations of super-Gaussian source priors:
a \emph{mixture} form, suited to MCMC, and an \emph{envelope} form, suited to
MAP optimisation. For a scalar $x$ and auxiliary variable $\lambda>0$,
\begin{align*}
p(x) &= \int p(x,\lambda)\,d\lambda \quad \text{(mixture)},\\
p(x) &= \sup_{\lambda>0} p(x,\lambda) \quad \text{(envelope)}.
\end{align*}
In our setting, $\lambda$ plays the role of a scale parameter in a normal
scale mixture, aligned with the auxiliary-variable framework of
\citet{geman1995nonlinear}.

We briefly recall the convex-analytic setup following \citet{polson2015mixtures}.
Let $\theta:\mathbb{R}^n\to\overline{\mathbb{R}}$ be a closed convex function
with convex conjugate
\[
\theta^\ast(\lambda) = \sup_{x\in\mathbb{R}^n}\{\lambda^\top x - \theta(x)\}.
\]
By the Fenchel--Moreau theorem,
\[
\theta(x)=\sup_{\lambda\in\mathbb{R}^n}\{\lambda^\top x-\theta^\ast(\lambda)\}.
\]

The key structural result is the following.

\begin{theorem}[\citet{polson2015mixtures}]
Let $p(x)\propto\exp\{-\phi(x)\}$ be symmetric in $x$, and define
$\theta(u)=\phi(\sqrt{2u})$ for $u>0$, with $\theta'(u)$ completely monotone.
Then $p(x)$ admits both a normal scale mixture and an envelope representation:
\[
e^{-\phi(x)}
\;\propto\;
\int_{\mathbb{R}^+} \NormRV(x\mid 0,\lambda^{-1})\,p_I(\lambda)\,d\lambda
\;\propto\;
\sup_{\lambda\ge0}
\bigl\{\NormRV(x\mid 0,\lambda^{-1})\,p_V(\lambda)\bigr\},
\]
where the \emph{variational prior} $p_V(\lambda)$ satisfies
\[
p_V(\lambda)\;\propto\;\lambda^{-1/2}\exp\{\theta^\ast(\lambda)\},
\]
and any optimal $\hat\lambda(x)$ lies in the subdifferential of
$\theta(x^2/2)$ or satisfies
\[
\hat\lambda(x) = \frac{\phi'(x)}{x}.
\]
\end{theorem}

The envelope representation induces a simple two-step iteration. For a generic
quadratic data term $(y-x)^2$ and penalty $\phi(x)$ we obtain
\[
x^{(t+1)} = \arg\min_x\bigl\{(y-x)^2 + \lambda^{(t)}x^2\bigr\},\qquad
\lambda^{(t+1)} = \frac{\phi'\bigl(x^{(t+1)}\bigr)}{x^{(t+1)}},\quad t=0,1,\dots
\]
This yields an EM/MM-type algorithm where the auxiliary $\lambda$ plays the
role of a local curvature parameter; large step-sizes are allowed early, and
the updates naturally adapt as the iterates approach a mode.

\paragraph{The $1/\cosh$ prior and P\'olya--Gamma mixtures.}
For the MacKay hyperbolic secant prior $p(s)\propto 1/\cosh(s)$, 
\citet{polson2015mixtures} show that
\begin{equation}
\frac{1}{\cosh(x)}
\;\propto\;
\int_0^\infty \NormRV(x\mid 0,\lambda^{-1})\,p_I(\lambda)\,d\lambda
\;=\;
\sup_{\lambda\ge0}\bigl\{\NormRV(x\mid 0,\lambda^{-1})\,p_V(\lambda)\bigr\},
\label{eq:cosh-mixture-envelope}
\end{equation}
where $p_I(\lambda)$ is proportional to the \PG$(1,0)$ density. Thus the
hyperbolic secant density admits both a P\'olya--Gamma normal scale mixture
representation \emph{and} an envelope representation. This links directly to
proximal and auxiliary-function algorithms: the normal mixture underpins
Gibbs or MH-within-Gibbs samplers, while the envelope form yields efficient
MAP updates via quadratic surrogates.

A concrete example is
\begin{align}
\log\cosh(x/2) 
&= \inf_{\lambda\ge0}\left\{\frac{\lambda}{2}x^2 - \theta^\ast(\lambda)\right\},\\
\hat\lambda(x) 
&= \E_{\text{PG}}(\lambda\mid x)
= \frac{1}{2x}\tanh\!\left(\frac{x}{2}\right),
\end{align}
using that $\log\cosh(\cdot)$ arises as a P\'olya--Gamma scale mixture.
In our ICA setting, the objective involves sums of $\log\cosh$ terms
\[
\sum_{n=1}^N\sum_{i=1}^d \log\cosh\bigl(w_i^\top \bm{x}^{(n)}\bigr),
\]
which can all be handled using the same envelope mechanism. For
statistical properties of $1/(\pi\cosh(x))$ as a density and $\log\cosh$
as a robust loss, see \citet{saleh2022statistical}.

\paragraph{Quadratic bounds.}
\citet{bouchard2007efficient} obtain a useful quadratic upper bound for
$\log\!\sum_i e^{x_i}$ in two steps. First, for any $\alpha\in\mathbb{R}$ and
$\bm{x}=(x_1,\dots,x_n)$,
\begin{equation}
\log\sum_{i=1}^n e^{x_i}
\;\le\;
\alpha + \sum_{i=1}^n \log\bigl(1+e^{x_i-\alpha}\bigr).
\label{eq:bou1}
\end{equation}
Second, they apply the standard tight quadratic bound for
$\log(1+e^x)$ \citep{jaakkola1997variational}:
\[
\log(1+e^x)
\;\le\;
\lambda(\xi)\,(x^2-\xi^2) + \frac{1}{2}(x-\xi) + \log(1+e^\xi),
\quad \forall\,\xi\in\mathbb{R},
\]
where $\lambda(\xi)= \frac{1}{2\xi}\bigl(\frac{1}{1+e^{-\xi}}-\frac{1}{2}\bigr)
= \frac{1}{4\xi}\tanh(\xi/2)$ and equality holds at $x=\xi$.
Combining this bound with \eqref{eq:bou1} yields a quadratic surrogate for
$\log\sum_i e^{x_i}$ in terms of auxiliary parameters $\{\xi_i\}$, which can
be updated in closed form. This is conceptually similar to the
P\'olya--Gamma mixture/envelope constructions we employ for $1/\cosh$.

\subsection{Auxiliary Function Based EM Algorithm}\label{subsec:aux-em}

We now derive an EM algorithm for the ICA unmixing matrix $\bm{W}$ under the
MacKay prior $p(s)\propto 1/\cosh(s)$ and show its equivalence to the
auxiliary-function optimisation of \citet{ono2010auxiliary}.

Let the sample log-likelihood for $\bm{W}$ (up to an additive constant) be
\begin{equation}\label{eq:obj-ica}
\ell(\bm{W})
= \log\text{det}\bm{W} - \sum_{i=1}^d \hat{E}\,\phi\bigl( \bm{w}_i^\top \bm{x}\bigr),
\quad
\phi(s)=\log\cosh(s),
\end{equation}
where $\bm{w}_i^\top$ is the $i$th row of $\bm{W}$ and $\hat{E}$ denotes the
empirical average over samples.

Using the P\'olya--Gamma scale mixture for $1/\cosh(\cdot)$, the complete-data
posterior for $(\bm{W},\bm{\lambda})$ (up to proportionality) is
\begin{align*}
p(\bm{W},\bm{\lambda}\mid \bm{x})
&\;\propto\;
\text{det}(\bm{W})^N
\prod_{n=1}^N \prod_{i=1}^d
\frac{1}{\cosh\bigl(\bm{w}_i^\top\bm{x}^{(n)}\bigr)}
\\
&\;\propto\;
\text{det}(\bm{W})^N
\prod_{n=1}^N \prod_{i=1}^d
\int_0^\infty
\exp\!\left\{-\frac{1}{2}\bigl(\bm{w}_i^\top\bm{x}^{(n)}\bigr)^2\lambda_{ni}\right\}
\sqrt{\lambda_{ni}}\,
p_I(\lambda_{ni})\,d\lambda_{ni},
\end{align*}
where $p_I(\lambda)$ is proportional to the \PG$(1,0)$ density.

\paragraph{E-step.}
The conditional expectation of each $\lambda_{ni}$ given $\bm{W}^{(t)}$ and
$\bm{x}^{(n)}$ is
\[
\hat{\lambda}_{ni}^{(t)}
=
\frac{\phi'\bigl(\bm{w}_i^{(t)\top}\bm{x}^{(n)}\bigr)}
{\bm{w}_i^{(t)\top}\bm{x}^{(n)}}
=
\frac{\tanh\bigl(\bm{w}_i^{(t)\top}\bm{x}^{(n)}\bigr)}
{\bm{w}_i^{(t)\top}\bm{x}^{(n)}}.
\]
The EM $Q$-function (per observation, up to constants) becomes
\begin{align*}
Q(\bm{W}\mid \bm{W}^{(t)})
&=
\log \text{det}(\bm{W})
- \sum_{i=1}^d
\hat{E}\left[
\frac{1}{2}\hat{\lambda}_{ni}^{(t)}
\bigl(\bm{w}_i^\top \bm{x}\bigr)^2
\right]\\
&=
\log \text{det}(\bm{W}) - \sum_{i=1}^d \bm{w}_i^\top Z_i^{(t)}\bm{w}_i,
\end{align*}
where
\[
Z_i^{(t)} :=
\hat{E}\left[
\frac{\tanh\bigl(\bm{w}_i^{(t)\top}\bm{x}\bigr)}
{2\,\bm{w}_i^{(t)\top}\bm{x}}\,\bm{x}\bm{x}^\top
\right]
=
\hat{E}\left[
\frac{\phi'\bigl(\bm{w}_i^{(t)\top}\bm{x}\bigr)}
{2\,\bm{w}_i^{(t)\top}\bm{x}}\,\bm{x}\bm{x}^\top
\right].
\]

\paragraph{M-step.}
Maximising $Q(\bm{W}\mid\bm{W}^{(t)})$ w.r.t.\ $\bm{W}$ is equivalent to
solving, for each row $\bm{w}_i$,
\[
\frac{\partial}{\partial \bm{w}_i}
Q(\bm{W}\mid\bm{W}^{(t)}) =
\bm{0},
\]
which yields the system
\begin{equation}
\bm{w}_j^\top Z_i^{(t)} \bm{w}_i =
\begin{cases}
1, & i=j,\\[2pt]
0, & i\neq j.
\end{cases}
\end{equation}
These equations are solved row-wise, subject to the implicit orthogonality
constraints induced by the log-determinant term. In practice, one can update
rows sequentially or use a joint Newton step constrained to $\text{GL}(d)$.

\paragraph{Equivalence to Ono's auxiliary function optimisation.}
\citet{ono2010auxiliary} construct an auxiliary function
$J(\bm{W},\bm{R})$ with slack variables $\bm{R}=(r_1,\dots,r_d)$ such that
\[
\ell(\bm{W}) = \max_{\bm{R}} J(\bm{W},\bm{R}),
\]
where
\[
J(\bm{W},\bm{R})
=
\log\text{det}\bm{W}
- \sum_{i=1}^d
\hat{E}\left[
\frac{\phi'(r_i)}{2r_i}\bigl(\bm{w}_i^\top\bm{x}\bigr)^2
+ F(r_i)
\right],
\quad
F(r) = \phi(r) - \frac{r\phi'(r)}{2},
\]
and $\phi(s)=\log\cosh(s)$ satisfies \citep[Theorem~1]{ono2010auxiliary}
\[
\phi(s)
\;\le\;
\frac{\phi'(r)}{2r}s^2 + F(r),\quad\forall r,
\]
with equality at $r=|s|$. Maximising $J(\bm{W},\bm{R})$ alternately over
$\bm{R}$ and $\bm{W}$ gives:

\begin{description}
\item[Update $\bm{R}$:]
Given $\bm{W}^{(t)}$, set
$r_i^{(t+1)} = \hat{E}\bigl|\bm{w}_i^{(t)\top}\bm{x}\bigr|$, which plugs back
into $J(\bm{W},\bm{R})$ and yields
\[
J(\bm{W},\bm{R}^{(t+1)})
=
\log\text{det}\bm{W} - \sum_{i=1}^d \bm{w}_i^\top Z_i^{(t)}\bm{w}_i + C^{(t)},
\]
for a constant $C^{(t)}$ independent of $\bm{W}$. This is exactly the EM
$Q$-function above, up to an additive constant.

\item[Update $\bm{W}$:]
Given $\bm{R}^{(t+1)}$, update
\[
\bm{W}^{(t+1)}
=
\arg\max_{\bm{W}}
J(\bm{W},\bm{R}^{(t+1)})
=
\arg\max_{\bm{W}}
\Bigl\{\log\text{det}\bm{W}
- \sum_{i=1}^d \bm{w}_i^\top Z_i^{(t)}\bm{w}_i\Bigr\},
\]
which coincides with the EM M-step.
\end{description}

Thus the auxiliary-function optimisation of \citet{ono2010auxiliary} is
precisely the latent-variable EM algorithm induced by the P\'olya--Gamma
normal scale mixture for the $1/\cosh$ prior, providing a clear Bayesian
interpretation and a direct bridge between our mixture-based MCMC and
envelope-based MAP procedures.

\subsection{Additional simulation study: EM versus MacKay's algorithm.}
We now compare the EM algorithm described above with MacKay's original
natural gradient method on synthetic data generated from a P\'olya--Gamma
scale-mixture model. Mimicking Section \ref{sec:toyexample}, we fix $N=500$, $d=4$, and consider
\begin{align}
v_{ij} &\sim \NormRV(0,\sigma_2^2), \quad
V\in\mathbb{R}^{d\times d}, \nonumber\\
\tau_{ni} &\sim PG(1,0), \quad
\tau\in\mathbb{R}^{N\times d}, \nonumber\\
s_{ni}\mid\tau_{ni}
&\sim \NormRV\bigl(0,(4\tau_{ni})^{-1}\bigr), \quad
S\in\mathbb{R}^{N\times d}, \nonumber\\
\bm{x}^{(n)}\mid\bm{s}^{(n)},V
&\sim \NormRV\bigl(V\bm{s}^{(n)},\sigma^2 I_d\bigr), \quad
X\in\mathbb{R}^{N\times d},
\label{eq:dgpnew}
\end{align}
so that, in matrix notation, $X = S V^\top + E$ with $E_{ni}\sim\NormRV(0,\sigma^2)$.

\begin{figure}[ht!]
    \centering
    \includegraphics[width=0.6\linewidth]{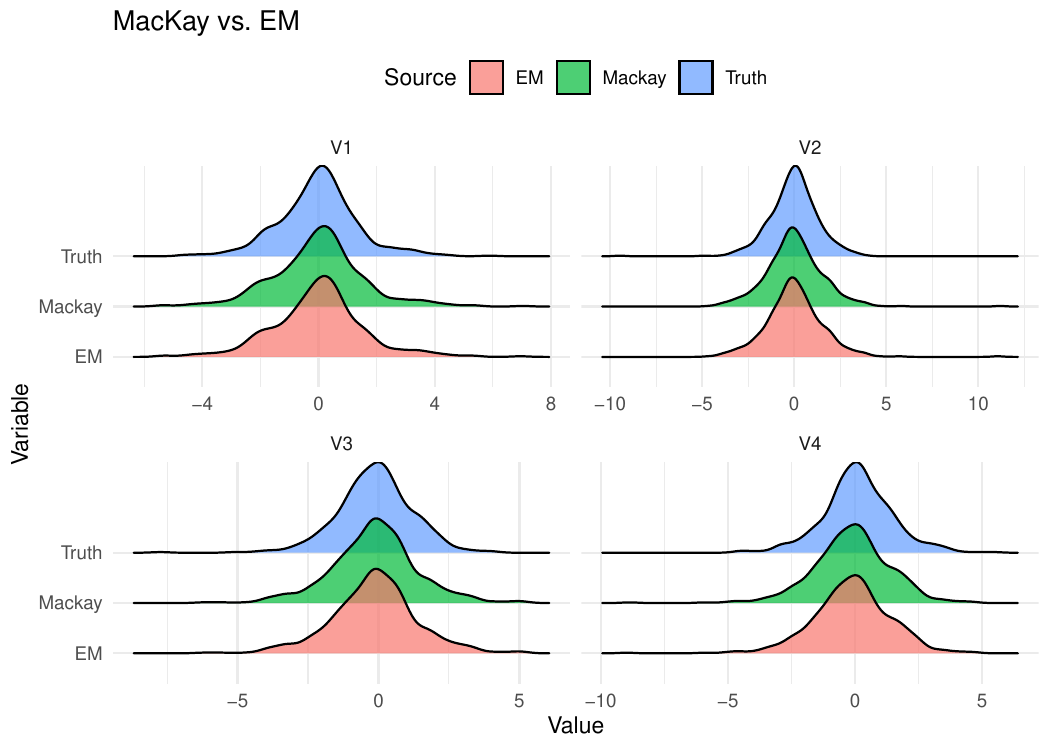}
    \caption{Comparison of the densities for $\hat{\bm{s}}$ and $\bm{s}$ for
    MacKay's algorithm and the EM algorithm under the data-generating process
    \eqref{eq:dgpnew} with $\sigma=0.01$ and $\sigma_2=1$.}
    \label{fig:ridge}
\end{figure}

\begin{figure}[ht!]
    \centering
    \includegraphics[width=0.6\linewidth]{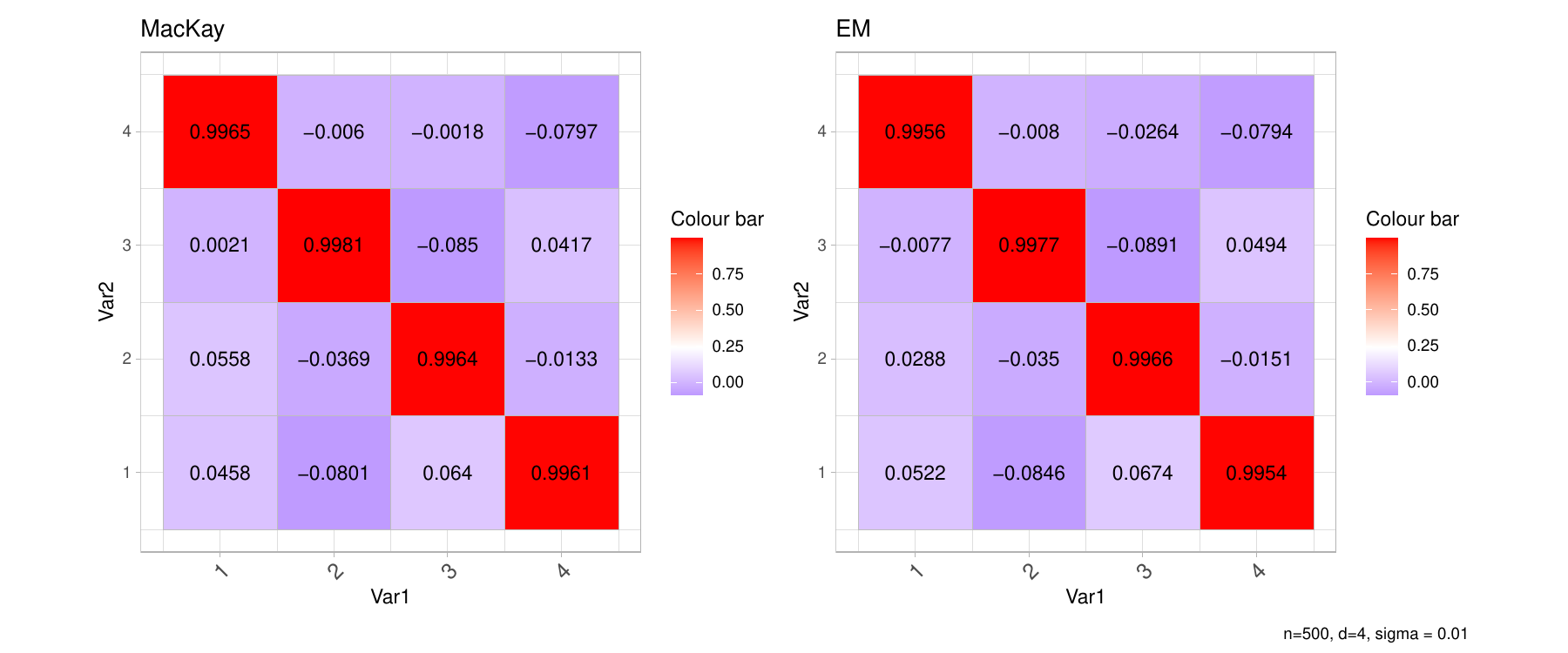}
    \caption{Correlations between $\hat{\bm{s}}$ and $\bm{s}$ for the two
    optimisation methods in the first experiment.}
    \label{fig:corr}
\end{figure}

We first set $\sigma=0.01$ and $\sigma_2=1$, and compare MacKay's original
algorithm with the EM algorithm derived in Section~\ref{subsec:aux-em}. The
estimated sources $\hat{\bm{s}}$ produced by each method are compared to the
true $\bm{s}$ via marginal density estimates (Figure~\ref{fig:ridge}) and
pairwise correlations (Figure~\ref{fig:corr}). In this regime, the two
algorithms exhibit very similar performance in terms of signal recovery: both
methods produce posterior modes that align well with the true sources, and
the empirical correlations between $\hat{\bm{s}}$ and $\bm{s}$ are close to
one for all components.

\begin{figure}[ht!]
    \centering
    \includegraphics[width=0.6\linewidth]{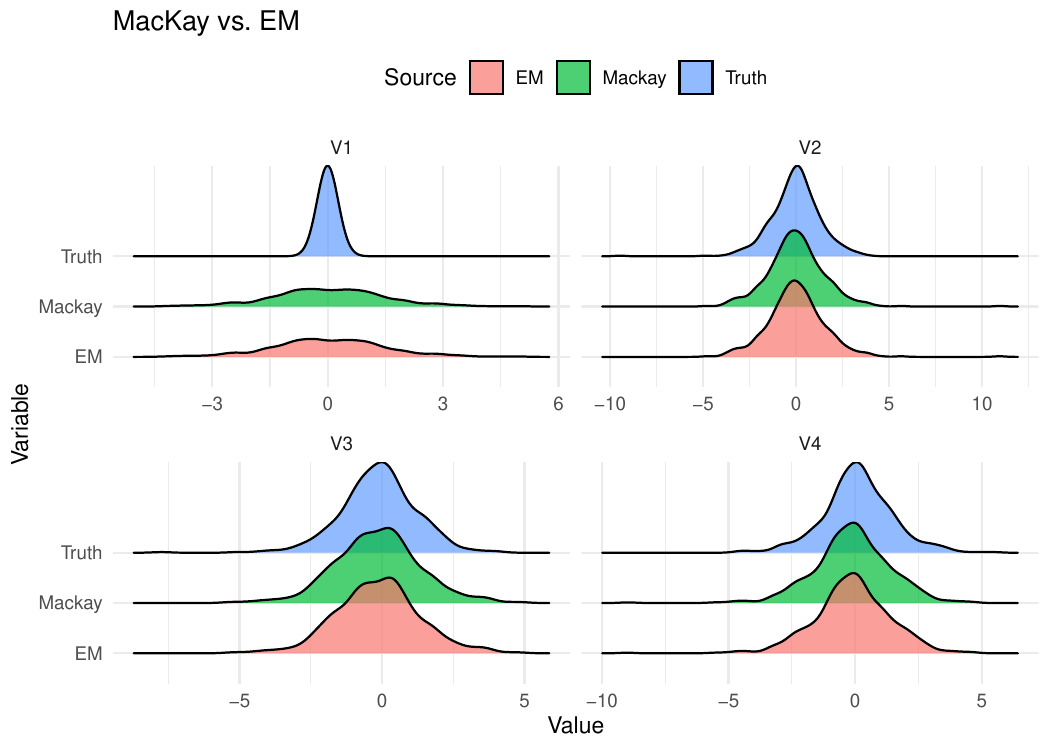}
    \caption{Comparison of the densities for $\hat{\bm{s}}$ and $\bm{s}$ after
    rescaling the first P\'olya--Gamma column by a factor of $100$ and setting
    $\sigma=0.1$ in \eqref{eq:dgpnew}. Both methods have difficulty in
    recovering the first source but perform similarly on the remaining
    components.}
    \label{fig:ridge2}
\end{figure}

\begin{figure}[ht!]
    \centering
    \includegraphics[width=0.6\linewidth]{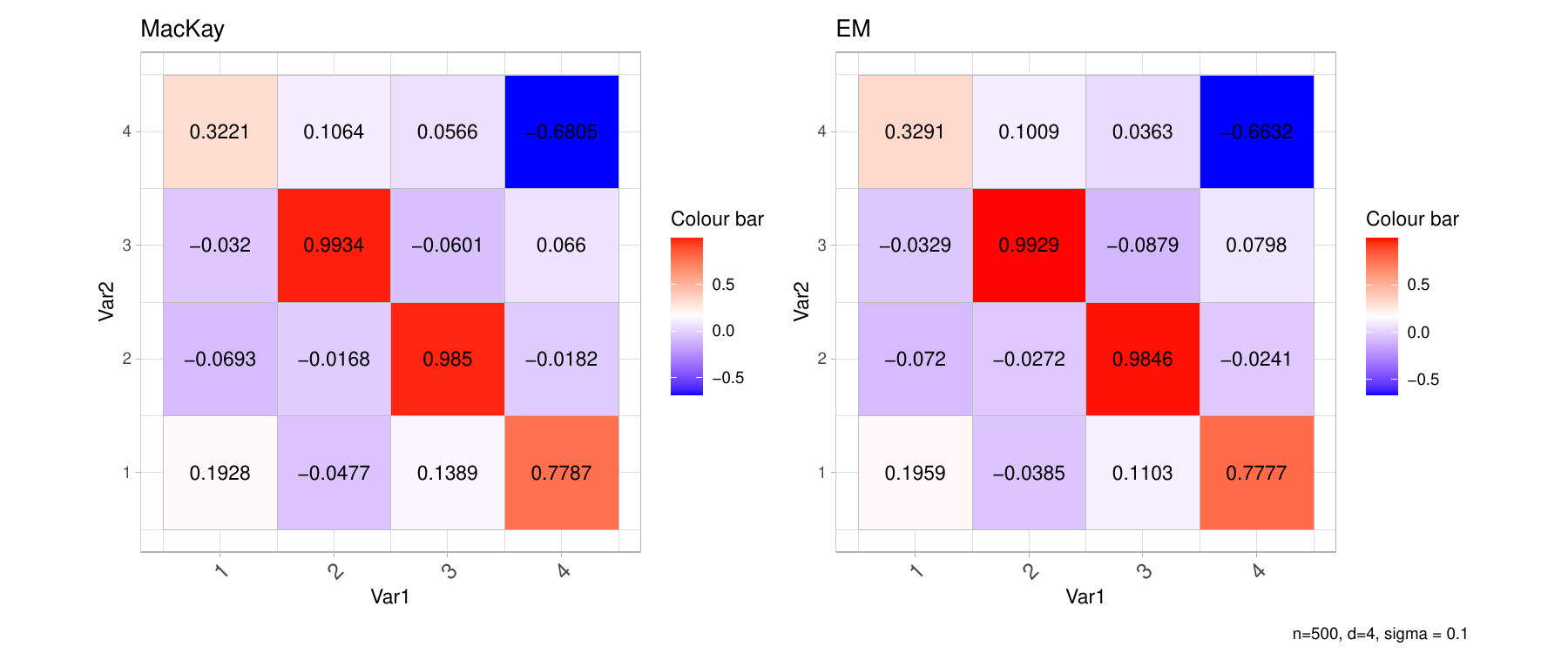}
    \caption{Correlations between $\hat{\bm{s}}$ and $\bm{s}$ for MacKay's
    algorithm and the EM algorithm in the second experiment.}
    \label{fig:corr2}
\end{figure}

In a second experiment, we increase the difficulty of the first component by
rescaling the corresponding P\'olya--Gamma variable by a factor of $100$ and
raising the noise level to $\sigma=0.1$, while keeping the rest of the setup
in \eqref{eq:dgpnew} unchanged. Concretely, we multiply the first column of
$\tau$ by $100$ before sampling $S$, which makes the first source direction
substantially harder to identify. We then rerun both MacKay's algorithm and
the EM algorithm. In this setting, both methods struggle to recover the first
signal but are still able to identify the remaining components, and again
their performance is very similar. Figures~\ref{fig:ridge2} and
\ref{fig:corr2} report the corresponding density estimates and correlation
plots.

\section{Normalizing Flows and Nonlinear ICA}
Recent advances in nonlinear independent component estimation have drawn heavily from the theory of normalizing flows and invertible neural networks. These models provide flexible bijective mappings between observed data and latent sources, offering a complementary perspective to the Bayesian super-Gaussian ICA framework developed in this paper. Below, we briefly review some of the key ideas from this line of work and clarify how flow-based architectures relate to the broader ICA literature.

\paragraph{NICE (non-linear independent component estimation)}
\citet{dinh2014nice} provide a deep learning framework called the  for high-dimensional density estimation, followed by the real NVP \citep{dinh2016density} transformations for unsupervised learning. The real NVP method learns a stable and invertible bijective function or map between samples $x \sim p_X$ and latent space $s \sim p_S$ or, $\theta \sim p_{\theta}$. For example, \citet{trippe2018conditional} utilize normalizing flows as likelihoods for conditional density estimation for complex densities. \citet{jimenez2015variational} provide an approximation framework using a series of parametric transformations for complex posteriors. A new method for Monte Carlo integration called Neural Importance Sampler was provided by \citet{muller2019neural} based on the NICE framework by parametrizing the proposal density by a collection of neural networks. 

\vspace{0.1in}

\noindent \textbf{Invertible Neural Network:} An important concept in the context of generative models is an invertible neural network or INNs \citep{dinh2016density}. Loosely speaking, an INN is a one-to-one function with a forward mapping $f: \mathbb{R}^d \mapsto \mathbb{R}^d$, and its inverse $g = f^{-1}$. \citet{song2019mintnet} provides the `MintNet' algorithm to construct INNs by using simple building blocks of triangular matrices, leading to efficient and exact Jacobian calculation. On the other hand, \citet{behrmann2021understanding} show that common INN method suffer from exploding inverses and provide conditions for stability of INNs. For image representation, \citet{jacobsen2018revnet} introduce a  deep invertible network, called the i-revnet, that retains all information from input data up until the final layer. 

\paragraph{Flow transformation models:} Here $\Y = h(\X) $ where $h(\cdot) $ is typically modeled as an invertible neural network (INN), with both $p_Y(\cdot)$ and $p_F(\cdot)$ as Gaussian densities, 
\begin{align*}
		p(\Y, \X \mid \S) &= p(\Y | \X, \S) p(\X | \S)\\
		&= p_y(\Y \mid h^{-1}(\X), \s) p_F(h^{-1}(\X)) \left \vert \frac{\partial h^{-1}}{\partial \X} \right \vert
\end{align*}
where the determinant of Jacobian is easy to compute. These models can be thought of as latent factor models. Flow-based methods can construct a nonlinear ICA where the dimensionality of the latent space is equal to the data as in an auto-encoder approach, see \citet{camuto2021towards}.


\noindent \textbf{Latent Factor Model:} Another interesting class of models contain latent factors that are driven with INNs.  These models take the form 
\begin{align}
		\Y &=  \F^{T} \S + \bepsilon\\
		\X &= h(\F)\\
		\F & \sim \NormRV(\0, \I_p),\; \S \sim p(\s)
\end{align}
where $(\Y, \X)$ are observed data and $\bepsilon$ is the mean zero Gaussian noise. Here $h(\cdot)$ is an invertible neural network (INN) and $\F$ are the latent factors.  This is essentially a flow transformation model and therefore, we can estimate $h$ and $\s$ using the loss function:
\[
L(h, \S) = \sum_{i=1}^N \left\{\lambda \Vert y_i - h^{-1}(x_i)^{T}\s\Vert^2 +  \Vert h^{-1}(x_i) \Vert^2 - \log \Big| \frac{\partial h^{-1}}{\partial x}\Big |(x_i) \right\} - \log p(\s)
\]
An iterative  two-step minimization procedure to learn $h$ and $\s$ is given by:
For $t=1,2,\ldots$
	\begin{enumerate}
		\item $\hat h^{(t)} = \hat h^{(t-1)} - \eta \nabla L(h, \hat\S^{(t-1)})$
		\item $\hat \S^{(t)} = \arg\min_\S L(\hat h^{(t)}, \S)$, or draw samples from the posterior $\propto \exp\left(-L(\hat h^{(t)}, \S)\right)$.
\end{enumerate}


\noindent \textbf{HINT (Hierarchical Invertible Neural Transport):} \citet{kruse2021hint, detommaso2019hint} provide the algorithm for posterior sampling. In this formulation, the function $T$ moves in the normalizing direction: a complicated and irregular data distribution $p_w(\w)$ towards the simpler, more regular or `normal' form, of the base measure $p_z(z)$. Let $\w := [\y, \x] \in \mathbb{R}^{m+d}$ and $T(\w):= [T^y(\y), T^x(\x, \y)] \in  \mathbb{R}^{m+d}$. The inverse function $S^{-1} = T$ is denoted as $S(\z) := [S^{y}(\z_y), S^{x}(\z_x, \z_y)]$ where $S^{\y} = (T^y)^{-1}$ where we assume that $\z \sim \NormRV(0, I_{m+d})$. As $p_z= p_{z_y} p_{z_x|z_y} $ and $S^{y}$ pushes forward the base density $p_{z_y}$ to $p_y$,  it can be shown that $S^x(\cdot, z_y)$ pushes forward the base density $p_{z_x|z_y}(\cdot | z_y)$ to the posterior density $p_{x|y}(\cdot | y)$, when $z_y = T^y(y)$. To sample from $p_{x|y}$, we simply sample $z_x \sim \NormRV(0, I_d)$ and calculate $x = S^{x}(z_x, z_y) = S^{x}(z_x, T^{y}(y))$, since $p_{z_x|z_y} = p_{z_x}$.

Two popular approaches for generative models that rely on latent space representation of the input data, albeit using substantially different architecture) are VAE (variational auto-encoders) and GAN (generative adversarial networks), but the iterative algorithms are only approximate based on a Kullback--Leiber divergence based approximation. On the theoretical side, \citet{wang2023data} provide exact proximal algorithms based on EM and MCMC algorithms. There are many directions for future work, particularly extensions to fields where traditional statistical methods dominate such as spatial or spatiotemporal data analysis. 

\end{document}